\documentclass[10pt,twocolumn]{IEEEtran}

\usepackage{amsmath,graphics,amssymb,epsfig,subfigure,color,cite}

\usepackage{amsthm}
\theoremstyle{plain}
\newtheoremstyle{mystyle}
  {0mm}
  {0mm}
  {}
  {4mm}
  {\bfseries}
  {:}
  { }
  {\thmname{#1}\thmnumber{ #2}\thmnote{ (#3)}}
\theoremstyle{mystyle}

\usepackage{array}
\usepackage{multirow}
\usepackage{enumerate}
\usepackage{bm}
\usepackage{float}
\usepackage{makecell}
\usepackage[right=0.625in, left=0.625in, top=0.7in, bottom=1.1in]{geometry}

\usepackage{algorithm2e}
\usepackage{algpseudocode}
\usepackage{algorithmicx}
\usepackage{algcompatible}
\usepackage{setspace}

\usepackage{bbm}
\usepackage{bigints}
\usepackage{adjustbox}

\floatname{algorithm}{Algorithm}

\usepackage{enumitem}
\newcommand{\argmax}{\operatornamewithlimits{argmax}}
\newcommand{\argmin}{\operatornamewithlimits{argmin}}
\makeatletter
\newcommand{\vast}{\bBigg@{4.5}}
\newcommand{\Vast}{\bBigg@{7.5}}
\makeatother

\begin{document}
\title{\fontsize{23}{28}\selectfont ESC-MVQ: End-to-End Semantic Communication With Multi-Codebook Vector Quantization
}

\author{Junyong Shin, Yongjeong Oh, Jinsung Park, Joohyuk Park, and Yo-Seb Jeon
	    \thanks{Junyong Shin, Jinsung Park, Joohyuk Park, Yongjeong Oh, and Yo-Seb Jeon are with the Department of Electrical Engineering, POSTECH, Pohang, Gyeongbuk 37673, Republic of Korea (e-mail: sjyong@postech.ac.kr; yongjeongoh@postech.ac.kr; jinsung@postech.ac.kr; joohyuk.park@postech.ac.kr; yoseb.jeon@postech.ac.kr).}
	}\vspace{-2mm}	
	
	\maketitle
	\vspace{-12mm}

\begin{abstract} 
This paper proposes a novel end-to-end digital semantic communication framework based on multi-codebook vector quantization (VQ), referred to as ESC-MVQ. Unlike prior approaches that rely on end-to-end training with a specific power or modulation scheme, often under a particular channel condition, ESC-MVQ models a channel transfer function as parallel binary symmetric channels (BSCs) with trainable bit-flip probabilities. Building on this model, ESC-MVQ jointly trains multiple VQ codebooks and their associated bit-flip probabilities with a single encoder-decoder pair. To maximize inference performance when deploying ESC-MVQ in digital communication systems, we devise an optimal communication strategy that jointly optimizes codebook assignment, adaptive modulation, and power allocation. To this end, we develop an iterative algorithm that selects the most suitable VQ codebook for semantic features and flexibly allocates power and modulation schemes across the transmitted symbols. Simulation results demonstrate that ESC-MVQ, using a single encoder-decoder pair, outperforms existing digital semantic communication methods in both performance and memory efficiency, offering a scalable and adaptive solution for realizing digital semantic communication in diverse channel conditions.
\end{abstract}

\begin{IEEEkeywords}
    Digital semantic communication, joint source-channel coding, end-to-end training, multi-codebook vector quantization, adaptive modulation and power allocation.
\end{IEEEkeywords}

\section{Introduction}\label{Sec:Intro}

Semantic communication is redefining the way information is transmitted by focusing on meaning rather than raw data accuracy. Unlike conventional systems that prioritize bit-level fidelity, semantic communication seeks to extract and convey only the most relevant information, ensuring that the received message aligns with its intended purpose. This approach is closely related to Weaver’s three levels of communication \cite{Weaver}, where the goal extends beyond technical correctness to achieving semantic understanding and effective information exchange. By operating at the semantic level, this paradigm offers a fundamental shift in communication design, particularly beneficial in extremely bandwidth-limited and power-constrained environments where traditional methods struggle to maintain efficiency \cite{Semantic_intro1,Semantic_intro2,Semantic_intro3}.

A major breakthrough in this field is the deep joint source-channel coding (DeepJSCC) \cite{DeepJSCC}, a deep learning-based joint source-channel coding approach. Unlike traditional separate source-channel coding schemes, DeepJSCC employs deep neural networks to extract semantic features from the source at the transmitter, while reconstructing the source data from perturbed features at the receiver to perform semantic tasks. By learning this process in an end-to-end manner, it achieves significant performance gains over the traditional separate coding schemes. Harnessing the power of deep learning, DeepJSCC has enabled highly efficient communication strategies, representing a key milestone in the advancement of semantic communication \cite{DeepJSCC, DeepJSCC-f, predJSCC}. The semantic features extracted by the deep neural network, however, are represented as a real-valued (e.g., 32-bit floating point) vector.
Transmitting such continuous-valued representations is either infeasible or highly inefficient in digital communication. Therefore, a direct implementation of the original DeepJSCC approach may not be compatible with modern digital communication systems. 

To realize semantic communications compatible with modern communication systems, digital semantic communication has been actively explored in recent studies. In \cite{constel_design, DeepJSCC-Q}, the semantic features were reformulated as complex numbers and then directly mapped into digital symbols. In this approach, a customized constellation was designed to suit the semantic features, enabling finite-rate transmission. 
Another approach for digital semantic communication is to employ a quantization module that converts semantic features into finite-length bit sequences. A commonly adopted method is scalar quantization (SQ), which independently quantizes each feature entry and then applies digital modulation to the resulting bits. For example, in \cite{BSC_Goldsmith, NECST}, a simple one-bit binarization was used to transform each feature entry into a binary bit. In \cite{Retrans.,sDMCM,BlindJSCC}, a multi-bit predefined uniform SQ was employed for the same purpose. However, digital semantic communication based on predefined SQ faces fundamental limitations. First, SQ is unable to exploit correlations among different feature entries, as it quantizes each entry independently. Moreover, the use of a fixed, predefined quantizer hinders effective adaptation to the specific statistical distributions of the semantic features. These limitations restrict the efficiency of finite-bit representations, ultimately degrading the overall system performance.

Recently, digital semantic communication based on vector quantization (VQ) has received increasing attention to establish a learnable and efficient finite-bit representation of the semantic features \cite{VQsem, Info_bottleneck, RobustVQVAE, SemTop, sDAC}.  The common idea behind these studies is to jointly train a learnable VQ codebook along with the encoder and decoder neural network to efficiently quantize its output features. This approach allows the quantizer to capture correlations among the feature entries and reflect their distribution, resulting in a more efficient finite-bit representation of the semantic feature with reduced quantization error \cite{VQ-VAE, shape-gain, ICTC, ECVQ, Wi-Fi}.
However, all these methods heavily rely on an end-to-end training for a specific communication environment (e.g., power/rate constraints or channel conditions), making them vulnerable to performance degradation when there is a mismatch between training and deployment scenarios. 

A straightforward approach to address the generalization issue is to utilize multiple models, each trained for a specific communication environment. However, this approach is not scalable to all possible scenarios, not only due to the vast diversity of communication environments but also because of the significant memory burden it imposes. To tackle this problem, an adaptive modulation scheme for the VQ outputs was introduced in \cite{UnivSC}, allowing the model to adjust its performance by selecting an appropriate modulation scheme based on the channel conditions. However, this method applies a fixed modulation order across all quantization outputs, which limits its flexibility in adapting to diverse and dynamic channel environments.
This limitation can be addressed by adopting a flexible modulation scheme, as proposed in \cite{Joohyuk}, or by incorporating a flexible power and modulation strategy, as explored in \cite{BlindJSCC}. However, these studies are restricted to the SQ framework and therefore suffer from non-learnable and inefficient finite-bit representations. A flexible power and modulation scheme for VQ-based digital semantic communication remains an open challenge. The key difficulty in this direction lies in the fact that the VQ codebook, trained under a specific power and modulation configuration, may not be compatible with other power or modulation schemes. To the best of the authors' knowledge, no existing studies have addressed how to enable flexible power and modulation for VQ-based digital semantic communication.


To fill this research gap, this paper proposes an end-to-end semantic communication framework with multi-codebook vector quantization,  referred to as ESC-MVQ. The framework is divided into two stages: a training stage and an inference stage. In the training stage, ESC-MVQ models a channel transfer function as parallel binary symmetric channels (BSCs) with trainable bit-flip probabilities, without assuming any fixed power, modulation scheme, or channel condition. Unlike prior approaches that rely on end-to-end training under a specific power or modulation scheme, often tailored to a particular channel condition, this design makes the trained VQ codebooks independent of specific communication scenarios. To further enhance flexibility in adapting to diverse channel conditions, we train multiple VQ codebooks and their associated bit-flip probabilities alongside a single encoder-decoder pair. In the inference stage, we aim to maximize the performance of ESC-MVQ in practical digital communication systems. To achieve this, we devise an optimal communication strategy that jointly assigns the most suitable VQ codebook to semantic features and flexibly allocates different power and modulation schemes across transmitted symbols. Simulation results demonstrate that the proposed ESC-MVQ framework outperforms existing digital semantic communication methods in terms of both performance and memory efficiency.

The main contributions of this paper are summarized as follows:
\begin{itemize}
\item For the training stage, we develop a novel end-to-end training strategy for jointly training multiple VQ codebooks with semantic encoder and decoder, enabling scalable and adaptive digital semantic communication. In this strategy, the transmission and reception process of binary outputs are modeled as parallel BSCs with trainable bit-flip probabilities. Based on this model, we jointly optimize multiple VQ codebooks and their associated bit-flip probabilities alongside a single semantic encoder-decoder pair, leveraging a carefully designed loss function that includes a regularization term. This regularization term controls the range of bit-flip probabilities, which are optimized to reflect the expected bit-error rate (BER) under various channel conditions.



\item For the inference stage, we devise optimal communication strategies for ESC-MVQ. Specifically, we formulate a joint codebook assignment, modulation, and power allocation problem under the constraint that the actual BERs in a digital communication system match the bit-flip probabilities learned for the VQ codebooks. To solve this problem, we develop a joint codebook assignment, adaptive modulation, and power allocation method. Additionally, we propose a low-complexity variant of this method by fixing the modulation order.

\item Through simulations, we demonstrate that the proposed ESC-MVQ framework outperforms existing digital semantic communication methods in an image reconstruction task. Furthermore, the proposed framework significantly improves memory efficiency by using a single encoder-decoder pair, in contrast to prior methods that require multiple pairs. These results collectively confirm that ESC-MVQ is a promising solution for realizing scalable and channel-adaptive digital semantic communication compatible with modern wireless standards.

\end{itemize}

This work builds upon our previous study \cite{Junyong_Globecom}, where we introduced only the training strategy for jointly training multiple VQ codebooks with semantic encoder and decoder. In the current work, we newly propose communication strategies to optimize the trained performance of ESC-MVQ under varying power constraints and channel conditions.
Based on the trained VQ codebooks and corresponding bit-flip probabilities, we jointly determine the codebook assignment for each sub-vector and optimize power and modulation allocation to minimize the distortion function.
Through this extension, our simulation results demonstrate the superiority of the proposed framework in enhancing the task performance and promising scalability in digital semantic communication systems.

    
\begin{figure*}[t]
    \centering
    {\epsfig{file=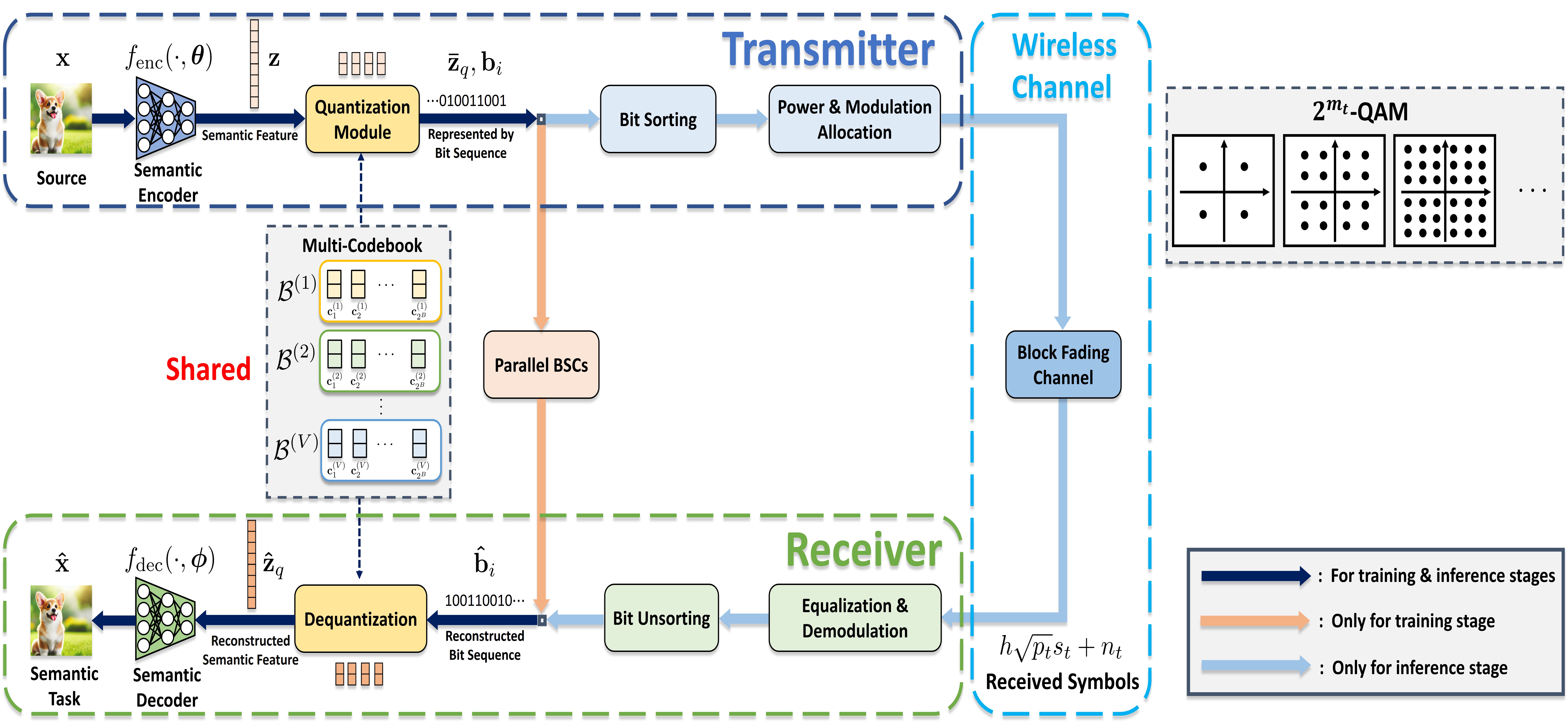,width=16cm}}
    \caption{An overall architecture of the proposed digital semantic communication framework.}\vspace{-3mm}
    
    \label{fig:System Model}
\end{figure*}


\section{System Model and Preliminary}\label{Sec:System_and_Preliminary}
In this section, we present a digital semantic communication system considered in the proposed framework.  

\subsection{Digital Semantic Communication System}\label{Sec:System}
A digital semantic communication system considered in this work is illustrated in Fig. \ref{fig:System Model}. To perform a specific semantic task with a given data sample $\mathbf{x}\in \mathbb{R}^{M_0}$, the transmitter first employs a semantic encoder network $f_{\rm enc}(\cdot,{\bm \theta})$ to extract a semantic feature $\mathbf{z}\in \mathbb{R}^{M'}$, given by
\begin{align}\label{eq: x-z}
\mathbf{z} = f_{\rm enc}(\mathbf{x}, {\bm \theta})\in \mathbb{R}^{M'}.
\end{align}
Since the elements of $\mathbf{z}$ are continuous-valued, they are quantized to enable finite-rate transmission via digital communication. A detailed description of the quantization process used in the proposed framework will be provided in the next subsection. 
After quantization, the bit sequence representing the quantized semantic feature is mapped to a symbol sequence $\mathbf{s}=[s_1,...,s_T]$ of length $T$ using digital modulation. The transmission rate, defined as the average number of bits carried per constellation symbol, is given by
\begin{align}\label{eq: transmission rate}
R = \frac{L}{T},
\end{align}
where $L$ denotes the bit sequence length. Let $p_t$ be the power allocated to the symbol $s_t$, where the total power constraint is given by 
\begin{align}\label{eq: power_tot}
    \sum^{T}_{t=1}p_t = P_{\rm tot}.
\end{align}
Then, each symbol $s_t$ is transmitted with power $p_t$ over a block-fading wireless channel, where the channel coefficients remain constant during the coherence time \cite{blockfading}. 


The received signal at time slot $t$ is given by
\begin{align}\label{eq: s-y}
y_t = h\sqrt{p_t}s_t + n_t,\ t\in\{1,...,T\},
\end{align}
where $h\in \mathbb{C}$ is the complex-valued channel coefficient, and $n_t$ represents the additive white Gaussian noise (AWGN) distributed as $n_t\sim \mathcal{CN}(0,\sigma^2)$. 
We assume that $h$ takes independent values when transmitting each semantic feature from different input data samples. Additionally, we assume that the channel coefficient is perfectly known at both the transmitter and receiver  through perfect channel estimation and feedback. Under these assumptions, the transmitted bit sequence is reconstructed at the receiver using a detection process. After detection, the reconstructed bit sequence is dequantized into $\hat{\mathbf{z}}_q$, which serves as the input to the semantic decoder $f_{\rm dec}(\cdot,{\bm \phi})$.
The receiver employs the semantic decoder as
\begin{align}\label{eq: x_hat}
    \mathbf{\hat x}=f_{\rm dec}(\hat{\mathbf{z}}_q,{\bm \phi}),
\end{align}
in order to perform the semantic task using the decoder output $\mathbf{\hat x}$.


\subsection{Product VQ Structure}\label{Sec:VQ-VAE}
    
\begin{figure}[t]
    \centering
    {\epsfig{file=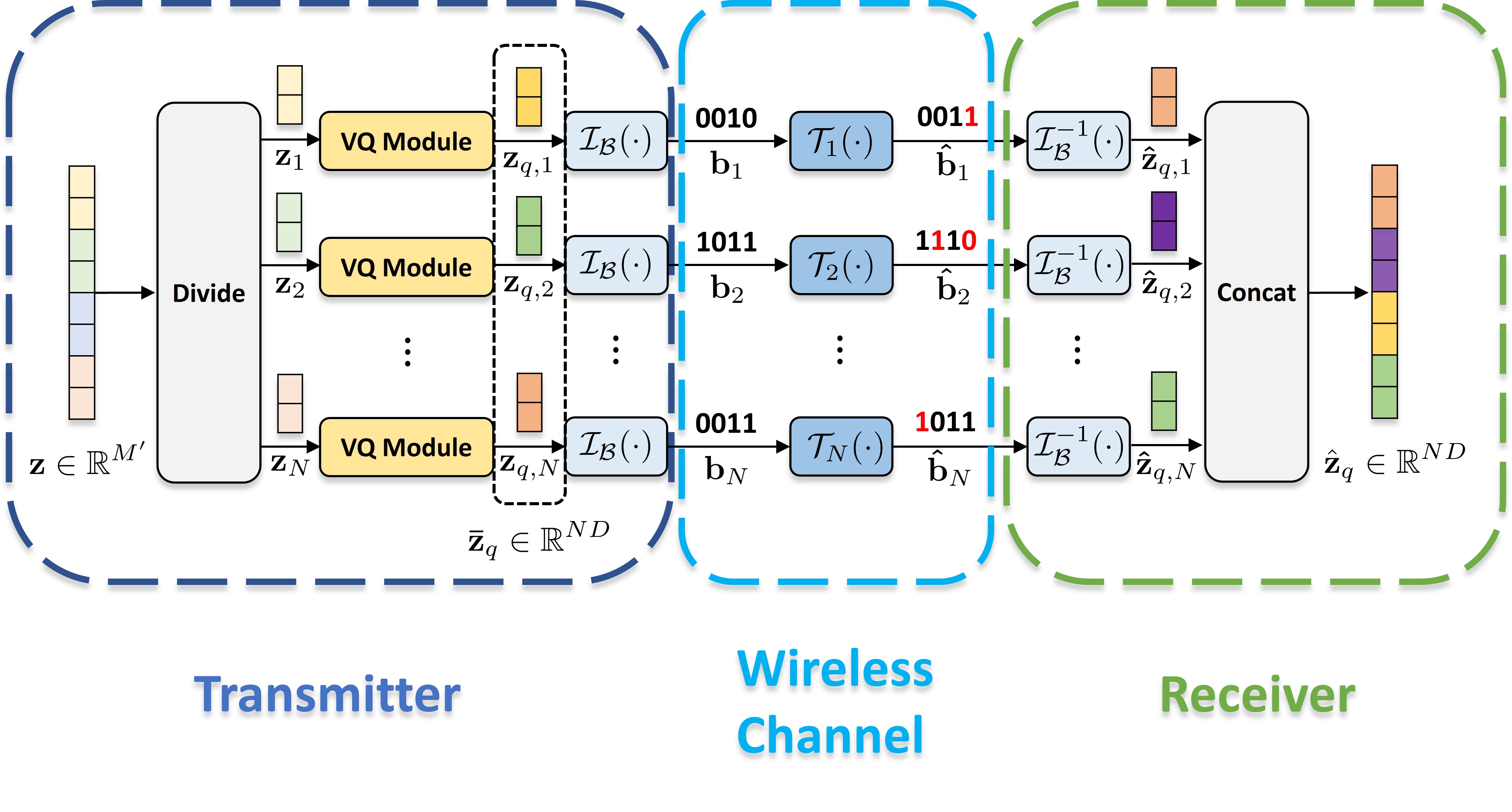,width=9cm}}
    \caption{An illustration of the operation of the proposed VQ module with a simplified model description. The transmitter applies product VQ and transmits binary sequences over the wireless channel. The receiver reconstructs the semantic feature from the received binary sequences.}\vspace{-3mm}
    \label{fig:VQVAE}
\end{figure}

In the digital communication system introduced above, it is crucial to employ an effective quantization method to mitigate the unwanted quantization errors between the original semantic feature $\mathbf{z}$ and its quantization $\bar{\bf z}_q$, as these errors may lead to a degradation in task performance at the receiver. 
A possible solution is to employ a predefined codebook, such as the Grassmannian codebook used in \cite{SemTop}. However, this approach cannot capture the specific distribution of the semantic feature $\mathbf{z}$ and thus may fail to achieve optimal performance. Since the distribution of semantic features depends on the task as well as the encoder and decoder networks, it is more beneficial to adopt a quantization method that can be tailored to this distribution. Motivated by this, we adopt a {\em learnable} VQ codebook that is jointly trained with the semantic encoder and decoder to maximize task performance. In particular, in this work, we consider a learnable product VQ structure as studied in \cite{productVQ}, which offers several advantages over standard SQ and VQ structures. First, compared to SQ, the use of a learnable VQ codebook provides a more flexible and efficient way to capture the distribution of semantic features. Second, compared to conventional VQ, the product VQ structure reduces memory requirements and computational complexity. The product VQ structure considered in our work is illustrated in Fig.~\ref{fig:VQVAE}. 

When employing the product VQ structure, the feature vector $\mathbf{z}$ is divided into $N$ sub-vectors, each of dimension $D$, and a $D$-dimensional codebook is used to quantize each sub-vector independently.
Consequently, the relationship $M'=N\times D$ holds.
Let $\mathcal{B}$ be a vector codebook using $B$ bits, consisting of $2^B$ codeword vectors, denoted as $\{\mathbf{c}_k\}^{2^B}_{k=1}$, each with dimension $D$.
For the $i$-th sub-vector ${\bf z}_i$ of ${\bf z}$, we define
\begin{align}\label{eq: z_i}
{\bf z}_i = [z_{(i-1)D+1},\cdots,z_{i D}]\in \mathbb{R}^D,
\end{align}
where $z_j$ refers to the $j$-th element of ${\bf z}$.
Each sub-vector ${\bf z}_i$ is quantized to $\mathbf{z}_{q,i}$ by selecting the nearest codeword $\mathbf{c}_k$ in $\mathcal{B}$ based on the Euclidean distance:
\begin{align}\label{eq: quantize}
    \mathbf{z}_{q,i}=\argmin_{\mathbf{c}_k \in \mathcal{B}}\Vert\mathbf{z}_i - \mathbf{c}_k \Vert^2.
\end{align}
 By concatenating the quantized sub-vectors, the quantized semantic feature is obtained as
\begin{align}
\bar{\bf z}_q =\big[\mathbf{z}_{q,1}^{\sf T},\cdots,\mathbf{z}_{q,N}^{\sf T}\big]^{\sf T}\in\mathbb{R}^{N D}.
\end{align}

For digital communication, each quantized sub-vector $\mathbf{z}_{q,i}$ is converted into the $B$-bit binary sequence:
\begin{align}\label{eq: bit index1}
    \mathbf{b}_{i} = {\mathcal I}_{\mathcal{B}}(\mathbf{z}_{q,i})\in \{0,1\}^{B},
\end{align}
where ${\mathcal I}_{\mathcal{B}}(\mathbf{z}_{q,i})$ is a binary mapper that outputs the binary representation of the index of the codeword vector $\mathbf{z}_{q,i}$. Consequently, the total length of the resulting bit sequence is calculated as $L=N\times B$.
After the digital communication process, described in the previous subsection, the reconstructed bit sequence can be represented as
\begin{align}
     \hat{\bf b}_i =  {\mathcal{T}}_i( {\bf b}_i )  \in \{0,1\}^{B},
\end{align}
where ${\mathcal{T}}_i({\bf b}_i)$ is the channel transfer function, which transforms the transmitted bit sequence ${\bf b}_i$ into the received binary sequence $\hat{\bf b}_i$ at the receiver as a result of the digital communication process. 
Then the reconstructed sub-vector is generated as follows:
\begin{align}
    \hat{\bf z}_{q,i} = \mathcal{I}_{\mathcal{B}}^{-1}(\hat{\bf b}_i) 
\end{align}
where ${\mathcal I}_{\mathcal{B}}^{-1}(\hat{\bf b}_i)$ is an inverse binary mapper which outputs the codeword vector in $\mathcal{B}$ whose index corresponds to $\hat{\bf b}_{i}$.
Utilizing the above notations, the reconstructed sub-vector can also be expressed as 
\begin{align}\label{eq:z_q_trans}
    \hat{\bf z}_{q,i} = {\mathcal{I}}_{\mathcal{B}}^{-1}({\mathcal{T}}_i( {\mathcal I}_{\mathcal{B}}(\mathbf{z}_{q,i})  )).
\end{align}
Then the reconstructed semantic feature is obtained as
\begin{align}
    \hat{\bf z}_{q} = \big[\hat{\bf z}_{q,1}^{\sf T},\cdots,\hat{\bf z}_{q,N}^{\sf T}\big]^{\sf T}\in\mathbb{R}^{N D}.
\end{align}
The relationship between $\hat{\bf z}_q$ and $\bar{\mathbf{z}}_{q}$ is illustrated in Fig.~\ref{fig:VQVAE}. Additionally, during model training, gradient correction is applied to the decoder input as follows:
\begin{align}
\mathbf{\hat z}_q \gets \mathbf{z} + \text{sg}(\mathbf{\hat z}_q - \mathbf{z})
\end{align}
where sg$(\cdot)$ denotes the ‘stop-gradient’ operator, which prevents its argument from participating in backpropagation, treating it as a constant.

\section{End-to-End Training Strategy for Semantic Communication with Multi-Codebook VQ}\label{Sec:Train}
In this section, we present the training stage of the proposed ESC-MVQ. We first introduce the BSC modeling as a generalized methodology for incorporating communication errors into the VQ module in the digital semantic communication. Next, we present a training strategy that optimizes the bit-flip probability in the BSC model through end-to-end learning. Finally, we propose a multi-codebook training strategy to enhance adaptability across diverse communication environments.

\subsection{Channel Transfer Function Modeling}\label{Sec:BSC}
When training the VQ module, it is essential to characterize the channel transfer function $\mathcal{T}_i(\cdot)$ in \eqref{eq:z_q_trans}, which determines the transformation from ${\bf z}_{q,i}$ to $\hat{\bf z}_{q,i}$. This function depends on the digital communication scenario, incorporating the effects of channel coefficients, noise power, power allocation, and modulation order, leading to diverse characteristics. A common approach to address this variability is to train separate models tailored to specific scenarios and select the appropriate model based on given conditions. However, this approach is impractical due to the excessive memory and computational complexity required to accommodate all possible digital communication scenarios.

To overcome this limitation, we consider parallel BSCs, providing a simple yet effective way to model communication errors in various digital communication scenarios. The BSC model has been utilized in training encoder and decoder networks within joint source-channel coding frameworks (e.g., \cite{NECST,BSC_Goldsmith}) and digital semantic communication frameworks (e.g., \cite{Joohyuk,BlindJSCC}). In our work, we incorporate the BSC model into the joint training of the VQ module with the digital semantic communication as in \cite{sDAC}.


 

Under the BSC modeling strategy, the channel transfer function $\mathcal{T}_i(\cdot)$ is modeled as $B$ parallel BSCs with a bit-flip probability sub-vector  ${\bm \mu}_i$, whose output follows the probability mass function given by 
\begin{align}\label{eq:transfer_function}
    \mathbb{P}\big(\mathcal{T}_{i}({\bf b}_i)= \hat{\bf b}_i \big)=\prod^{B}_{j=1}{\mu_{i,j}}^{\mathcal{H}_{i,j}}(1-\mu_{i,j})^{1-\mathcal{H}_{i,j}},
\end{align}
where $\mu_{i,j}$ is the $j$-th element of ${\bm \mu}_i$, and $\mathcal{H}_{i,j}$ denotes the Hamming distance between $[{\bf b}_i]_j$ and $[\hat{\bf b}_i]_j$, returning 1 if the bits differ and 0 otherwise. By concatenating the bit-flip probability sub-vectors, we define a bit-flip probability vector as
\begin{align}
    \bar{\bm \mu} =\big[{\bm \mu}_{1}^{\sf T},\cdots,{\bm \mu}_{N}^{\sf T}\big]^{\sf T}.
\end{align}
It is noteworthy that $\mu_{i,j}$ mimics the bit error probability when transmitting the $j$-th bit of the index of the codeword vector ${\bf z}_{q,i}$ through digital communication. This will be further clarified in Sec.~\ref{Sec:infer}.

By incorporating the channel transfer function in \eqref{eq:transfer_function} into \eqref{eq:z_q_trans}, the transition probability from the quantized sub-vector ${\bf z}_{q,i}$ to the reconstructed sub-vector $\hat{\bf z}_{q,i}$ is given by  
\begin{align}\label{eq: transition}
    p (\hat{\bf z}_{q,i}|{\bf z}_{q,i}) &\overset{(a)}{=} p (\hat{\bf b}_{i}|{\bf b}_{i})  = \mathbb{P}\big(\mathcal{T}_{i}({\bf b}_i)= \hat{\bf b}_i\big) \nonumber \\
    &=\prod^{B}_{j=1}{\mu_{i,j}}^{\mathcal{H}_{i,j}}(1-\mu_{i,j})^{1-\mathcal{H}_{i,j}},
\end{align}
where ${(a)}$ holds because both the binary mapper $\mathcal{I}_{\mathcal{B}}(\cdot)$ and the inverse binary mapper $\mathcal{I}_{\mathcal{B}}^{-1}(\cdot)$ are deterministic one-to-one mapping functions.
By modeling the channel transfer function through the BSCs, communication errors can be effectively captured during the training process of the VQ module.  
\subsection{End-to-End Training for ESC-MVQ}\label{Sec:Blind}
Based on the channel transfer function modeling in Sec.~\ref{Sec:BSC}, we present an end-to-end training strategy for the ESC-MVQ, which allows the joint optimization of the semantic encoder ${\bm \theta}$, the VQ codebook $\mathcal{B}$, and semantic decoder ${\bm \phi}$ along with the bit-flip probability vector $\bar{\bm \mu}$.  
By treating $\bar{\bm \mu}$ as a learnable parameter, our approach enables end-to-end optimization that captures the varying importance of individual bits, rather than relying on fixed or uniformly sampled bit-flip probabilities. This formulation eliminates the need for manual tuning and allows the model to adaptively allocate power and modulation based on semantic relevance. As a result, it leads to improved task performance compared to conventional uniform allocation strategies, as will be demonstrated in Sec.~V.
To this end, we consider a joint loss function given by 
\begin{align}\label{eq: VQ-VAE Loss}
\mathcal{L}_{\text{vq}} = d_0(\mathbf{x}, \mathbf{\hat x})+d_1(\mathbf z, \hat{\bf z}_q)
\end{align}
where $d_0(\mathbf{x}, \mathbf{\hat x})$ is a task-dependent loss term (e.g., mean squared error; MSE) and $d_1(\mathbf z, \hat{\bf z}_q)$ is a communication loss term.
The expression in \eqref{eq: VQ-VAE Loss} clearly demonstrates that our loss function takes into account both the task performance, captured by $d_0(\mathbf{x}, \mathbf{\hat x})$, and communication error, captured by $d_1(\mathbf z, \hat{\bf z}_q)$.
In particular, the second loss term considers not only the quantization error introduced by the VQ codebook $\mathcal{B}$ but also the communication error caused by the channel transfer function, which is absorbed into $\hat{\bf z}_q$.

Under our channel transfer function modeling, direct end-to-end training is not feasible because the transition from ${\bf z}_{q,i}$ to $\hat{\bf z}_{q,i}$ in \eqref{eq: transition} involves probabilistic sampling, which prevents direct backpropagation.
To address this challenge, we employ the Gumbel-Softmax trick \cite{GumbelSoftmax}, which enables differentiable sampling from a categorical distribution. Specifically, it represents the sampled outcome as a deterministic function of $\bar{\bm \mu}$, allowing backpropagation.
Based on the Gumbel-Softmax trick, the reconstructed sub-vector $\hat{\bf z}_{q,i}$ is expressed as 
\begin{gather}
    { \hat{\bf z}_{q,i} = \sum^{2^B}_{k=1}e_{i,k}\mathbf{c}_k,}
\end{gather}
where
\begin{gather}\label{eq: gumbel_softmax}
    e_{i,k}=\frac{{\rm exp}\{({\rm log}~p(\mathbf{c}_k|\mathbf{z}_{q,i})+g_{i,k})/\tau\}}{\sum^{2^B}_{u=1}{\rm exp}\{({\rm log}~p(\mathbf{c}_u|\mathbf{z}_{q,i})+g_{i,u})/\tau\}},
\end{gather}
and $g_{i,k}$ is a random variable independently drawn from a $\textit{Gumbel}(0,1)$ distribution. The parameter $\tau$ regularizes the softness of the Softmax function in the Gumbel-Softmax trick. As $\tau$ is annealed to zero during training, the trick gradually approximates a hard selection of the VQ codeword vector. Notably, the conventional use of the Gumbel-Softmax trick for the VQ module has been limited to efficient codebook updates, as in \cite{Info_bottleneck}. In contrast, our work aims to leverage this trick for training the bit-flip probability vector $\bar{\bm \mu}$ of the BSCs alongside the VQ module.  

Based on the Gumbel-Softmax trick, the bit-flip probability vector $\bar{\bm \mu}$ can be jointly trained with the semantic encoder, VQ codebook, and semantic decoder. However, as training progresses, all elements of $\bar{\bm \mu}$ continue to decrease because minimizing the bit-flip probabilities consistently reduces the loss function in \eqref{eq: VQ-VAE Loss}.  In particular, the optimal solution of $\bar{\bm \mu}$ under the loss $\mathcal{L}_{\rm vq}$ is zero, as this loss includes the communication error term $d_1(\mathbf z, \hat{\bf z}_q)$. Consequently, this causes the BSC model to diverge from practical communication scenarios, where bit errors cannot be arbitrarily reduced due to channel fading and noise. 
To prevent the bit-flip probabilities from becoming arbitrarily small, we modify our loss function by introducing a regularization term:
\begin{align}\label{eq: blvq}
    \mathcal{L}_{\text{esc-vq}} = \mathcal{L}_{\rm vq} +\lambda \mathcal{R}(\bar{\bm \mu}),
\end{align}
where $\mathcal{R}(\bar{\bm \mu})$ is a regularization function, defined as 
\begin{align}\label{eq: regularization}
    \mathcal{R}(\bar{\bm \mu})=\frac{1}{NB}\sum_{i} \sum_j \mu_{i,j}{\rm log}\mu_{i,j},
\end{align}
and $\lambda$ is a hyperparameter that determines the relative importance of $\mathcal{R}(\bar{\bm \mu})$. 
 In \cite{BlindJSCC}, where the bit-flip probability $\mu_{i,j}$ is learned within a fixed SQ framework, an MSE-based regularization function $\mathcal{R}(\bar{\bm \mu})$ is employed to penalize deviations from $0.5$. However, this approach tends to polarize the bit-flip probabilities toward either $0$ or $0.5$, thereby limiting the capability to capture diverse BERs across different power levels. To address this issue, we select the regularization function defined in \eqref{eq: regularization}, whose minimum is achieved when $\mu_{i,j}$ approaches a specific probability value, namely $\frac{1}{e}\approx 0.3679$. Consequently, a smaller $\lambda$ places more emphasis on minimizing $\mathcal{L}_{\mathrm{vq}}$, encouraging the bit-flip probabilities to decrease, whereas a larger $\lambda$ emphasizes the regularization term $\mathcal{R}(\bar{\bm \mu})$, driving the probabilities to increase. Thus, the bit-flip probabilities can be effectively controlled by appropriately selecting the hyperparameter $\lambda$.
\subsection{Multi-Codebook Training for ESC-MVQ}\label{Sec:Multi-Codebook}
The end-to-end training strategy in Sec.~\ref{Sec:Blind} enables the semantic encoder, VQ codebook, and decoder to be robust against the bit-flip probabilities of the parallel BSCs. However, the bit-flip probabilities obtained from end-to-end training may not always be achievable in practical digital communication scenarios. For instance, when the channel gain is low or the transmission power is highly constrained, it may be impossible to achieve actual bit-error probabilities in digital communication that are lower than the trained bit-flip probabilities. Thus, relying on a single set of bit-flip probabilities may limit the model’s ability to accommodate diverse digital communication scenarios.

To enhance flexibility and adaptability in end-to-end training, we propose a multi-codebook training strategy that jointly optimizes multiple VQ codebooks, each tailored to support a distinct set of bit-flip probabilities. This approach enables the transmitter to dynamically select the most suitable VQ codebook based on the achievable bit-flip probabilities, thereby improving adaptability across diverse communication scenarios.
In our strategy, we train $V$ distinct VQ codebooks along with their associated bit-flip probability vectors under different regularization conditions. This is achieved by assigning $V$ different values of the regularization parameter $\lambda$, denoted as $\lambda^{(1)} < \cdots < \lambda^{(V)}$, to the loss function in \eqref{eq: blvq}. Additionally, we impose different constraints on the bit-flip probabilities by ensuring that the bit-flip probabilities for the $v$-th VQ codebook remain above a certain minimum value, denoted as $\mu_{\rm min}^{(v)}$, where $\mu_{\rm min}^{(1)} < \cdots < \mu_{\rm min}^{(V)}$. This minimum value, together with $\lambda^{(v)}$, influences the distribution of bit-flip probabilities. During training, any $\mu^{(v)}_{i,j}$ falling below $\mu^{(v)}_{\min}$ is clipped to $\mu^{(v)}_{\min}$, and this clipping operation does not affect the gradient computation. Let $\mathcal{B}^{(v)}=\{\mathbf{c}^{(v)}_k\}^{2^B}_{k=1}$ be the $v$-th VQ codebook, and let $\bar{\bm \mu}^{(v)} \in [\mu_{\rm min}^{(v)},0.5]^{NB}$ represent the bit-flip probability vector associated with $\mathcal{B}^{(v)}$. The $i$-th quantized sub-vector using $\mathcal{B}^{(v)}$ is then determined as
\begin{align}
    \mathbf{z}_{q,i}^{(v)}=\argmin_{\mathbf{c}_k \in \mathcal{B}^{(v)}}\Vert\mathbf{z}_i - \mathbf{c}_k \Vert^2.
\end{align}
Similarly, the transition probability $p(\hat{\bf z}^{(v)}_{q,i}|\mathbf{z}^{(v)}_{q,i})$ is expressed as 
\begin{align}\label{eq: transition_multi}
    p (\hat{\bf z}_{q,i}^{(v)}|{\bf z}_{q,i}^{(v)}) =\prod^{B}_{j=1}\big({\mu_{i,j}^{(v)}}\big)^{\mathcal{H}_{i,j}^{(v)}}\big(1-\mu_{i,j}^{(v)}\big)^{1-\mathcal{H}_{i,j}^{(v)}},
\end{align}
where $\mu_{i,j}^{(v)}$ is the $j$-th element of ${\bm \mu}_i^{(v)}$ and $\mathcal{H}_{i,j}^{(v)}$ denotes the Hamming distance between the $j$-th elements of $\mathcal{I}_{\mathcal{B}^{(v)}}({\bf z}_{q,i}^{(v)})$ and $\mathcal{I}_{\mathcal{B}^{(v)}}(\hat{\bf z}_{q,i}^{(v)})$.
Based on the Gumbel-Softmax trick, the reconstructed sub-vector $\hat{\bf z}^{(v)}_{q,i}$ is expressed as
\begin{gather}\label{eq: gumbel_softmax_multi}
    \hat{\bf z}^{(v)}_{q,i} = \sum^{2^B}_{k=1}e^{(v)}_{i,k}\mathbf{c}^{(v)}_k,
\end{gather}
where
\begin{gather}
    e^{(v)}_{i,k}=\frac{{\rm exp}\{({\rm log}~p(\mathbf{c}^{(v)}_k|\mathbf{z}^{(v)}_{q,i})+g^{(v)}_{i,k})/\tau\}}{\sum^{2^B}_{u=1}{\rm exp}\{({\rm log}~p(\mathbf{c}^{(v)}_u|\mathbf{z}^{(v)}_{q,i})+g^{(v)}_{i,u})/\tau\}},
\end{gather}
and $g^{(v)}_{i,k}\sim\textit{Gumbel}(0,1)$.

{\small \RestyleAlgo{ruled}
    \SetKwComment{Comment}{/* }{ */}
    {\small \begin{algorithm}[t]\label{alg: training}
    \caption{Training strategy for ESC-MVQ}
    {\bf 1. Initialization:}\\
    Randomly initialize ${\bm \theta},{\bm \phi}, \mathcal{B}^{(1)}$, and $\mathcal{M}^{(1)}$;\\
    {\bf 2. Training:}\\
    \For{$v\in\{1,...,V\}$}{
    \While{not converged}{
    $\mathbf{z} = f_{\text{enc}}(\mathbf{x},{\bm \theta})$;\\
    \For{$u\in\{1,...,v\}$}{
    ${\mathbf{z}^{(u)}_{q,i} = {\argmin}_{\mathbf{b}_k\in\mathcal{B}^{(u)}} \Vert\mathbf{z}_i-\mathbf{b}_k \Vert^2}, \forall i$; \\
                
    $\mathbf{\tilde z}^{(u)}_{q,i} =$  $\sum^{2^B}_{k=1}e^{(u)}_{i,k}\mathbf{b}^{(u)}_k$ in \eqref{eq: gumbel_softmax_multi}, $\forall i$; \\
                
    $\mathbf{\tilde z}^{(u)}_q\leftarrow\mathbf{z}+\text{sg}(\mathbf{\tilde z}^{(u)}_q-\mathbf{z})$;\\ 
            
    $\mathbf{\hat x}^{(u)} = f_{\text{dec}}(\mathbf{\tilde z}_q^{(u)},{\bm \phi})$;\\
    }
    Calculate $\mathcal{L}^{(v)}_{\text{esc-mvq}}$ in \eqref{eq: BLVQ Loss-multi3} and update parameters;\\
    }
    $\mathcal{B}^{(v+1)} = \mathcal{B}^{(v)}$;\\
    $\mathcal{M}^{(v+1)} = \mathcal{M}^{(v)}$;
    }
    \end{algorithm}}}

A straightforward approach to training $V$ distinct VQ codebooks is to use $V$ separate encoder-decoder pairs, with each pair trained to support a specific VQ codebook. In this case, the end-to-end training can be applied independently to each encoder-decoder pair. However, this approach is inefficient, as it requires storing multiple models, leading to significant memory overhead. To address this challenge, we employ a single encoder-decoder pair and  develop a sequential multi-codebook training strategy to train it.

Our strategy consists of $V$ sequential training steps where the $v$-th step is designed to simultaneously train the first $v$ VQ codebooks.
More specifically, at the start of the $v$-th step, the $v$-th codebook $\mathcal{B}^{(v)}$ and the corresponding bit-flip probability vector $\bar{\bm \mu}^{(v)}$ are initialized using the previously trained parameters $\mathcal{B}^{(v-1)}$ and $\bar{\bm \mu}^{(v-1)}$, respectively. Then, we jointly train the $v$ codebooks $\mathcal{B}^{(1)}, . . . , \mathcal{B}^{(v)}$ and bit-flip probability vectors $\bar{\bm \mu}^{(1)}, . . . , \bar{\bm \mu}^{(v)}$ using the loss function $\mathcal{L}^{(v)}_{\text{esc-mvq}}$ defined as
\begin{align}
\mathcal{L}^{(v)}_{\text{esc-mvq}} &= \frac{1}{\sum_{k=1}^{v}\eta^k}\sum_{u=1}^{v}\eta^u \mathcal{L}^{(u)}_{\text{esc-vq}}, \label{eq: BLVQ Loss-multi3} \\
\mathcal{L}^{(u)}_{\text{esc-vq}} &= \mathcal{L}_{\text{vq}}^{(u)} +\lambda^{(u)}\mathcal{R}(\mathcal{M}^{(u)}), \label{eq: BLVQ Loss-multi2}  \\
\mathcal{L}_{\text{vq}}^{(u)} &= d_0(\mathbf{x},\mathbf{\hat x}^{(u)})+d_1(\mathbf{z},\hat{\bf z}_q^{(u)}), \label{eq: BLVQ Loss-multi1} 
\end{align}
where $\mathbf{\hat x}^{(u)}=f_{\rm dec}(\hat{\bf z}_q^{(u)},{\bm \phi})$ represents the decoder output from $\hat{\bf z}_q^{(u)}$, and $\eta\leq 1$ is a hyperparameter that regulates the contribution of each codebook's loss in $\mathcal{L}^{(v)}_{\text{esc-mvq}}$. Our multi-codebook training strategy enables joint optimization of the encoder and decoder while simultaneously training multiple VQ codebooks, each tailored to support a distinct set of bit-flip probabilities. The overall training procedure is summarized in Algorithm \ref{alg: training}.

\begin{figure}[t]
    \centering
    {\epsfig{file=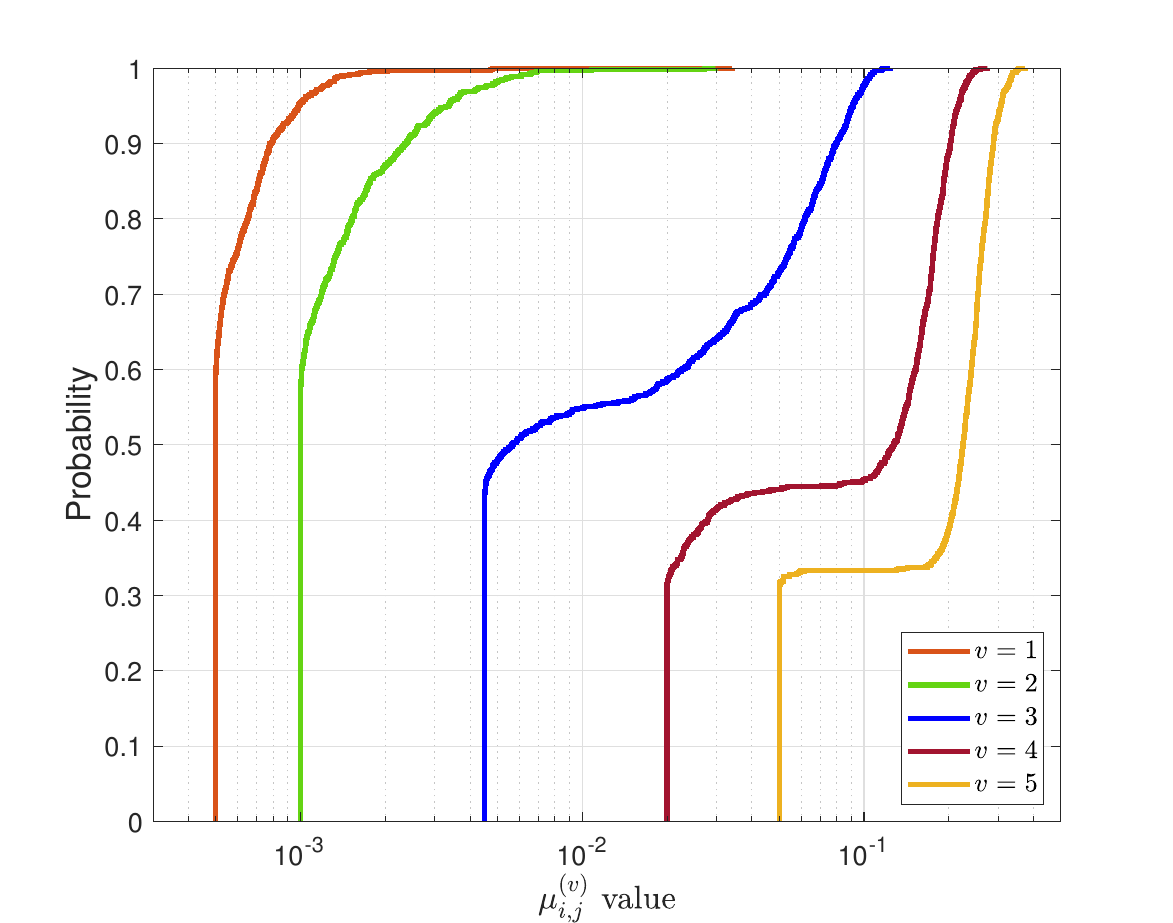,width=6cm}}
    \caption{The CDF of bit-flip probabilities $\mu^{(v)}_{i,j}$ trained with our multi-codebook training strategy using the CIFAR-100 dataset.}\vspace{-3mm}
    
    \label{fig:mu_cdf}
\end{figure}

{\bf Numerical Example (Distribution of Bit-Flip Probabilities):} We present an example illustrating the distribution of bit-flip probabilities after training with our multi-codebook training strategy. Fig. \ref{fig:mu_cdf} shows the cumulative distribution function (CDF)  of $\mu^{(v)}_{i,j}$ trained with our multi-codebook strategy using the CIFAR-100 dataset. The relevant parameters are set as $V=5$, $[\mu^{(1)}_{\rm min},...,\mu^{(5)}_{\rm min}]=[0.0005,0.001,0.0045,0.02,0.05]$, and $\lambda^{(v)}=\frac{1}{8}\cdot 2^{(v-1)},\ \forall v\in\{1,...,5\}$. Fig. \ref{fig:mu_cdf} illustrates that a VQ codebook trained with a larger value of $\lambda^{(v)}$ and $\mu^{(v)}_{\rm min}$ result in higher bit-flip probabilities, as expected. Additionally, Fig. \ref{fig:mu_cdf} confirms that the bit-flip probabilities associated with different VQ codebooks follow distinct distributions. These results highlight that our multi-codebook training strategy effectively enhances the flexibility and adaptability of the trained semantic encoder and decoder.

\section{Optimal Communication Strategy for ESC-MVQ}\label{Sec:infer}  
In our ESC-MVQ framework, the multi-codebook training strategy offers $V$ distinct VQ codebooks, each trained to support a specific set of bit-flip probabilities. This design enables the transmitter, during the inference stage, to dynamically assign the most suitable VQ codebook to each sub-vector, considering both the quality of the codebook and the feasibility of achieving the corresponding bit-flip probabilities. To maximize inference performance when deploying ESC-MVQ in practical digital semantic communication systems, this section presents an optimal communication strategy that jointly optimizes codebook assignment, adaptive modulation, and power allocation. 

    
\subsection{Problem Formulation}
We formulate an optimization problem to jointly optimize VQ codebook, modulation order, and transmit power assignment for ESC-MVQ.  
Let $v_i$ be the index of the VQ codebook assigned to the $i$-th quantized sub-vector ${\bf z}_{q,i}$. 
The codebook assignment for $N$ sub-vectors is then represented by a vector ${\bf v}=[v_1,\cdots,v_N] \in\{1,\ldots,V\}^{N}$, which we refer to as the {\em codebook assignment} vector.
To formulate the objective function of the optimization problem, we quantify the distortion associated with the codebook assignment. 
To this end, we define a distortion function of the $i$-th sub-vector using the $v_i$-th VQ codebook as
\begin{align}\label{eq: distortion}
    \mathcal{D}_i(v_i) 
    &= \mathbb{E}_{\hat{\bf z}^{(v_i)}_{q,i}|\mathbf{z}_i} [\|\hat{\bf z}^{(v_i)}_{q,i} - \mathbf{z}_i\|^2] \nonumber \\
    &\overset{(a)}{=} \sum^{2^B}_{k=1}p(\mathbf{c}^{(v_i)}_k|\mathbf{z}^{(v_i)}_{q,i})\cdot \big\Vert \mathbf{c}^{(v_i)}_k - \mathbf{z}_i \big\Vert^2,
\end{align}
where $(a)$ is computed from the transition probability in \eqref{eq: transition_multi}. 
This function measures the mean squared error between  the $i$-th sub-vector $\mathbf{z}_i$ and the corresponding received sub-vector $\hat{\bf z}^{(v_i)}_{q,i}$ with the $v_i$-th VQ codebook.
Given the power constraint and channel coefficients, the values of $v_i$ are determined by solving an optimization problem that minimizes the distortion function in \eqref{eq: distortion} at both the transmitter and receiver. However, a key challenge in this setup is that the receiver cannot directly access the true values of the sub-vectors $\mathbf{z}_i$ required for computing the distortion in \eqref{eq: distortion}. To address this challenge, we approximate the exact distortion using its empirical average over the entire training dataset: 
\begin{align}\label{eq: residual_approx}
\mathcal{D}_i(v_i)  \approx 
\underbrace{\frac{1}{|\mathcal{D}_{\rm train}|}\sum_{{\bf x} \in\mathcal{D}_{\rm train}} \mathcal{D}_i(v_i)\big|_{\mathbf{z}=f_{\rm enc}(\mathbf{x},{\bm \theta})}}_{\triangleq \hat{\mathcal{D}}_i(v_i)},
\end{align}
where $\mathcal{D}_{\rm train}$ denotes the training dataset. The approximated values are stored as a precomputed $V\times N$ table at both transmitter and receiver. This table is then used to solve the optimization problem.

When employing our multi-codebook training strategy in Sec. \ref{Sec:Multi-Codebook}, the $v_i$-th VQ codebook is trained to support parallel BSCs with the bit-flip probability vector ${\bm \mu}_{i}^{(v_i)}$ for the $i$-th sub-vector. 
Thus, the overall bit-flip probability vector associated with the codebook assignment vector ${\bf v}$ is given by
\begin{align}\label{eq:Bit_Flip}
    \tilde{\bm \mu}({\bf v}) =\big[({\bm \mu}_{1}^{(v_1)})^{\sf T},\cdots,({\bm \mu}_{N}^{(v_N)})^{\sf T}\big]^{\sf T}.
\end{align}
To maximize the performance of ESC-MVQ, we must ensure that the actual BERs in a digital communication system match the bit-flip probabilities in $\tilde{\bm \mu}({\bf v})$.
For an uncoded QAM system, where the bit sequence representing the quantized semantic feature $\bar{\bf z}_q$ is mapped to a symbol sequence $\mathbf{s}=[s_1,...,s_T]$, 
the BER for a symbol $s_t$ modulated with $m_t$-QAM and transmit power $p_t$ is approximated as  \cite{BER_approx}:
\begin{align}\label{eq: BER approx}
    {\epsilon}(p_t;m_t) &\approx \frac{\sqrt{2^{m_t}}-1}{\sqrt{2^{m_t}}\log_2\sqrt{2^{m_t}}}{\rm erfc}\left(\sqrt{\frac{3p_t\gamma}{2(2^{m_t}-1)}}\right) \nonumber\\
    &~~~+\frac{\sqrt{2^{m_t}}-2}{\sqrt{2^{m_t}}\log_2\sqrt{2^{m_t}}}{\rm erfc}\left(3\sqrt{\frac{3p_t\gamma}{2(2^{m_t}-1)}}\right) \nonumber \\
    &\triangleq \hat{\epsilon}(p_t;m_t),
\end{align}
where $\gamma  \triangleq \frac{|h|^2}{\sigma^2}$ is the channel-gain-to-noise-power ratio, and ${\rm erfc}(x)\triangleq1-\frac{2}{\sqrt{\pi}}\int_0^{x}{\rm exp}(-u^2)\ {\rm d}u$ is the complementary error function. 
This expression suggests that the bit-flip probabilities associated with the symbol $s_t$ can be aligned with the actual BER in \eqref{eq: BER approx} by adjusting the transmit power $p_t$ and modulation order $m_t$ according to the channel gain $|h|^2$. 
Based on this result, the BER-matching condition\footnote{It is also possible to adopt a more sophisticated BER-matching condition. For example, in $m_t$-QAM, the $m_t$ bits grouped into a single symbol may experience different BERs depending on their positions; thus, one could match each bit to its corresponding BER individually. Nevertheless, for simplicity, in this work, we assume that all bits within a symbol share the same average BER.}  is given by 
\begin{align}\label{eq:BER_match}
    \hat{\epsilon}(p_t;m_t) = \frac{1}{m_t}\sum_{u=\bar{m}^{(t-1)}+1}^{\bar{m}^{(t-1)}+m_t} {\sf sort}_u(\tilde{\bm \mu}({\bf v})), 
\end{align}
where $\bar{m}^{(t-1)} = \sum_{i=1}^{t-1}m_i$, with $\bar{m}^{(0)}=0$ and ${\sf sort}_u({\bf a})$ denotes the $u$-th smallest element of a vector ${\bf a}$. For example, if $\tilde{\bm \mu}({\bf v})=[0.1, 0.01, 0.02, 0.09]$, then ${\sf sort}_2(\tilde{\bm \mu}({\bf v}))=0.02$, which is the second smallest value in $\tilde{\bm \mu}({\bf v})$. To clarify the sorting operation in \eqref{eq:BER_match}, suppose $m_1=2$. In this case, the bits with bit-flip probabilities 0.01 and 0.02 are grouped as a symbol, and power is then allocated to match BER of $0.015$, the average of the two bit-flip probabilities. The sorting operation ensures that bits with similar bit-flip probabilities are modulated using the same symbol, improving the consistency of BER matching. Here, $\hat{\epsilon}(p_t;m_t)$ represents the BER induced by the actual communication channel while $\tilde{\bm \mu}({\bf v})$ represents the trained bit-flip probabilities within a model. Eventually, $\tilde{\bm \mu}({\bf v})$ is used to derive appropriate power and modulation settings through BER matching. Note that our ESC-MVQ matches the bit-flip probabilities trained over parallel BSCs to the BERs of practical modulation schemes, eliminating the need for retraining or fine-tuning for specific modulation types or orders. In other words, ESC-MVQ operates as a generalized semantic communication strategy that is not tied to any particular modulation scheme.

Based on the distortion function and BER-matching condition, the joint optimization problem for ESC-MVQ is formulated as
\begin{subequations}\label{eq:Opt_0}
\begin{align}
    \min_{{\bf v},{\bf m},{\bf p}}~&  \sum_{i=1}^{N}\hat{\mathcal{D}}_i(v_i), \\
    \text{s.t.}~~&\hat{\epsilon}(p_t;m_t) = \frac{1}{m_t}\sum_{u=\bar{m}^{(t-1)}+1}^{\bar{m}^{(t-1)}+m_t} {\sf sort}_u(\tilde{\bm \mu}({\bf v})),~\forall t,
    \label{eq:Opt_0_C1} \\
    & \sum_{t=1}^T p_t \leq  P_{\rm tot},~p_t\geq 0, \label{eq:Opt_0_C2}\\
    &\frac{1}{T}\sum_{t=1}^T m_t \geq R,~m_t \in\{2,4,...,m_{\rm max}\},\label{eq:Opt_0_C3}
\end{align}
\end{subequations}
where ${\bf p}=[p_1,\cdots,p_T]^{\sf T}$ is a power allocation vector, and ${\bf m}=[m_1,\cdots,m_T]^{\sf T}$ is a modulation order vector. 
This problem aims to minimize the overall distortion while optimizing power allocation, modulation order, and codebook assignment, subject to a total power constraint.
Our optimization problem is combinatorial and involves non-convex constraints, making it computationally challenging to solve, especially for a large number of sub-vectors and VQ codebooks.
To address this challenge, we develop two practical algorithms to efficiently solve the optimization problem in \eqref{eq:Opt_0}, which will be introduced in the following subsections.

\begin{figure}[t]
    \centering
    {\epsfig{file=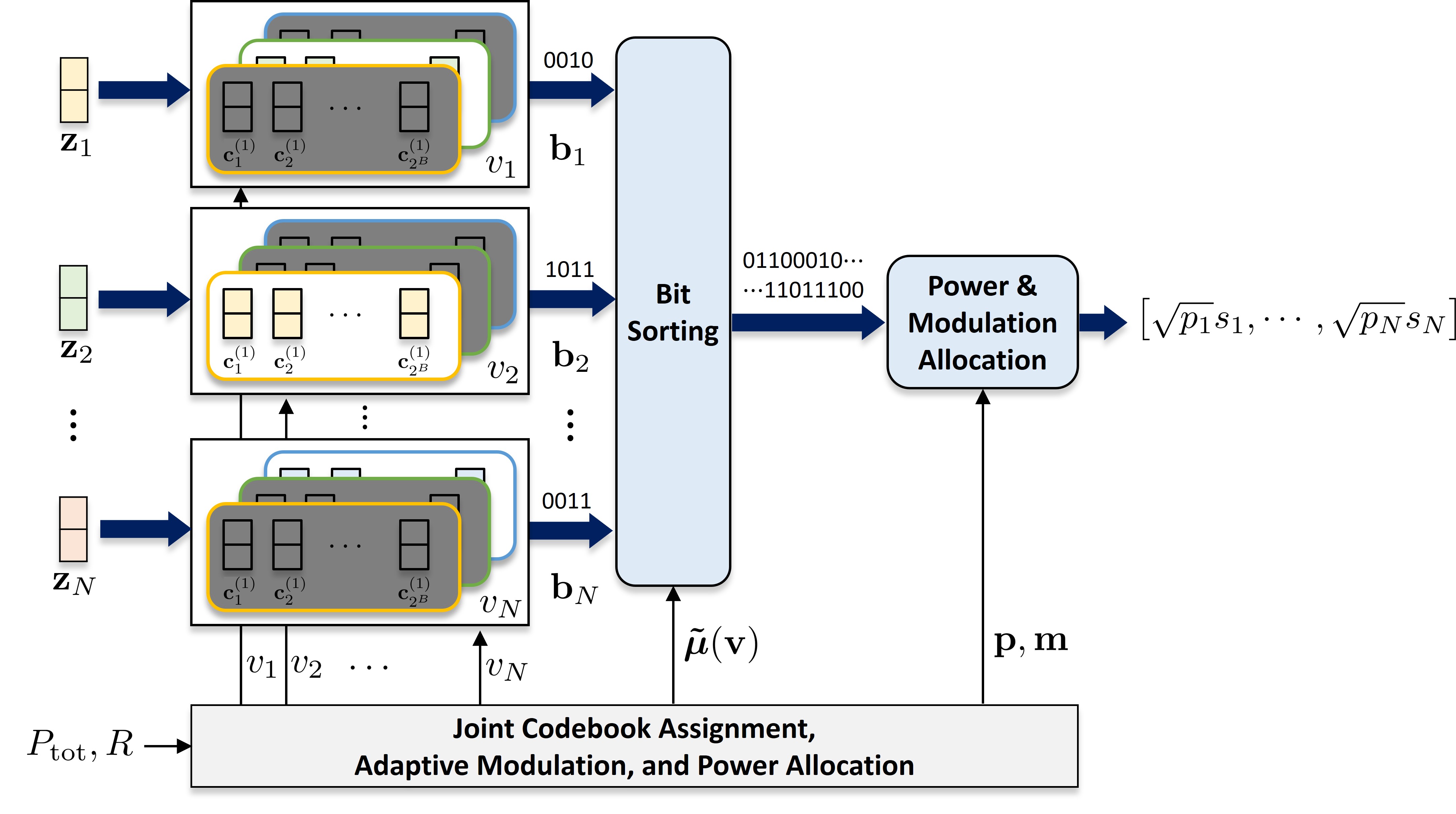,width=8.5cm}}
    \caption{An illustration of the JCAMP method.}\vspace{-3mm}
    
    \label{fig:JCAMP}
\end{figure}

\subsection{Joint Codebook Assignment, Adaptive Modulation, and Power Allocation (JCAMP) Method}\label{Sec:JCAMP}
To address the difficulties in solving the optimization problem \eqref{eq:Opt_0}, we present a joint codebook assignment, adaptive modulation, and power allocation (JCAMP) method, which is illustrated in Fig. \ref{fig:JCAMP}.
A major challenge arises from the fact that $m_t$ bits are entangled in the BER-matching constraint in \eqref{eq:Opt_0_C1} as these bits are jointly modulated as a single symbol. 
To circumvent this challenge, we temporarily assume that the BER-matching constraint is given in a per-bit basis as follows:
\begin{align}\label{eq:BER_match_temp}
    \hat{\epsilon}(\hat{p}_{i,j}\hat{m}_{i,j};\hat{m}_{i,j}) = \mu_{i,j}^{(v_i)},
\end{align}
where $\hat{p}_{i,j}$ and $\hat{m}_{i,j}$ is defined as a temporary power and temporary modulation order, respectively. 
Then our optimization problem \eqref{eq:Opt_0} can be reformulated as 
\begin{subequations}\label{eq:Opt_1}
\begin{align}
    \min_{{\bf v},\hat{\bf M},\hat{\bf P}}~&  \sum_{i=1}^{N}\hat{\mathcal{D}}_i(v_i), \\
    \text{s.t.}~~&\hat{\epsilon}(\hat{p}_{i,j}\hat{m}_{i,j};\hat{m}_{i,j}) = \mu_{i,j}^{(v_i)},~\forall i,j,
    \label{eq:Opt_1_C1} \\
    &\sum_{i=1}^N \sum_{j=1}^B \hat{p}_{i,j} \leq  P_{\rm tot},~ \hat{p}_{i,j}\geq 0.\\
    &\frac{1}{NB}\sum_{i=1}^N \sum_{j=1}^B \hat{m}_{i,j} \geq R, ~\hat{m}_{i,j} \in  \{2,4,...,m_{\rm max}\},
\end{align}
\end{subequations}
where $\hat{\bf M}$ and $\hat{\bf P}$ are the temporary power and temporary modulation order matrix, respectively, such that $[\hat{\bf M}]_{i,j} =\hat{m}_{i,j}$ and $[\hat{\bf P}]_{i,j} =\hat{p}_{i,j}$.
As the BER function $\hat{\epsilon}(\hat{p}_{i,j}\hat{m}_{i,j},\hat{m}_{i,j})$ is monotonically decreasing with respect to $\hat{p}_{i,j}$, the temporary power can be efficiently determined by solving the equation \eqref{eq: BER approx} for $\hat{p}_{i,j}$ using numerical methods such as bisection or Newton-Raphson algorithms \cite{p_alg.}. We define the obtained solution as the optimal temporary power denoted as 
\begin{align}
  \hat{p}_{i,j}^\star  = \frac{\hat{\epsilon}^{-1}(\mu_{i,j}^{(v_i)} ; \hat{m}_{i,j})}{\hat{m}_{i,j}}.
\end{align}
Then the problem simplifies to 
\begin{subequations}\label{eq:Opt_1-2}
\begin{align}
    \min_{{\bf v},\hat{\bf M}}~&  \sum_{i=1}^{N}\hat{\mathcal{D}}_i(v_i), \\
    \text{s.t.}~~&
    \sum_{i=1}^N \sum_{j=1}^B \hat{p}_{i,j}^\star  \leq  P_{\rm tot},~ \hat{p}_{i,j}^\star  = \frac{\hat{\epsilon}^{-1}(\mu_{i,j}^{(v_i)} ; \hat{m}_{i,j})}{\hat{m}_{i,j}}, \\
    &\frac{1}{NB}\sum_{i=1}^N \sum_{j=1}^B \hat{m}_{i,j} \geq R, ~\hat{m}_{i,j} \in  \{2,4,...,m_{\rm max}\},
\end{align}
\end{subequations}
To solve the optimization problem \eqref{eq:Opt_1-2}, we employ an alternating optimization approach which alternatively optimizes the temporary modulation order matrix $\hat{\bf M}$ and codebook assignment vector ${\bf v}$. To this end, we decompose the optimization problem into two sub-problems, where the first sub-problem optimizes the codebook assignment vector ${\bf v}$ for a fixed $\hat{\bf M}$, while the second sub-problem optimizes the temporary modulation order matrix $\hat{\bf M}$ for a fixed ${\bf v}$.

Our first sub-problem aims to minimize the overall distortion by optimizing the codebook assignment vector ${\bf v}$ for a fixed $\hat{\bf M}$ as follows:
\begin{subequations}\label{eq:Opt_1-3}
\begin{align}
    \text{(P1)}~\min_{{\bf v}} ~&\sum_{i=1}^{N}\hat{\mathcal{D}}_i(v_i), \\ 
    \text{s.t.}~&\sum_{i=1}^N \sum_{j=1}^B \hat{p}_{i,j}^\star  \leq  P_{\rm tot},~ \hat{p}_{i,j}^\star  = \frac{\hat{\epsilon}^{-1}(\mu_{i,j}^{(v_i)} ; \hat{m}_{i,j})}{\hat{m}_{i,j}}. \label{eq:Opt_1-3_C1}
\end{align}
\end{subequations}
To solve this problem in an efficient manner, we devise a greedy allocation algorithm, inspired by \cite{discrete optimization}, that gradually minimizes the overall distortion $\sum_{i=1}^{N}\hat{\mathcal{D}}_i(v_i)$, while considering the power consumption to satisfy the constraint in \eqref{eq:Opt_1-3_C1}. In the proposed greedy algorithm, we initialize the codebook assignment vector by choosing the highest codebook index (i.e., $v_i=V,\ \forall i$). This ensures that every sub-vector initially uses the lowest power. During iterations, we gradually reduce certain $v_i$ values step-by-step to meet the power constraint. In particular, at each iteration, we select the index $i^*$ that maximizes the distortion reduction relative to power increase:
\begin{align}\label{eq:JCAMP_criterion}
    &i^*=\nonumber\\ 
    &\argmax_{i:v_i>1} \frac{\hat{\mathcal{D}}_i(v_i) -\hat{\mathcal{D}}_i(v_i-1)}{\sum_{j=1}^B \big\{ \hat{\epsilon}^{-1}(\mu_{i,j}^{(v_i-1)} ; \hat{m}_{i,j}) - \hat{\epsilon}^{-1}(\mu_{i,j}^{(v_i)} ; \hat{m}_{i,j})\big\}/\hat{m}_{i,j}},
\end{align}

This criterion ensures that we prioritize changes that maximize distortion reduction while minimizing additional power consumption.

{ \RestyleAlgo{ruled}
    \SetKwComment{Comment}{/* }{ */}
    {\small \begin{algorithm}[t]\label{alg:JCAMP}
    \caption{Proposed JCAMP method}
    $v_i = V$, $\hat{m}_{i,j} = \frac{NB}{T}$, $\forall i,j$;\\
    \While{$\sum_{i,j}\hat{\epsilon}^{-1}(\mu_{i,j}^{(v_i)} ; \hat{m}_{i,j})/\hat{m}_{i,j} \leq {P}_{\rm tot}$}{
    \textbf{// Optimization for }(P1):\\
    \While{$\sum_{i,j}\hat{\epsilon}^{-1}(\mu_{i,j}^{(v_i)} ; \hat{m}_{i,j})/\hat{m}_{i,j} \leq {P}_{\rm tot}$}{
    $\hat{m}_{i,j}^{\rm buffer} = \hat{m}_{i,j}$, $\forall i,j$;\\
    Choose $i^*$ from \eqref{eq:JCAMP_criterion};\\
    $v_{i^*} \leftarrow v_{i^*}-1$;\\}
    \textbf{// Optimization for }(P2):\\
    $\hat{m}_{i,j} = \hat{m}_{i,j}^{\rm buffer}$, $\forall i,j$;\\
    \For{$m \in\{4,..., m_{\rm max}-2\}$}{
    \While{\textbf{True}}{
    Determine $p^+_{i,j}$ from \eqref{eq: p+} and $p^-_{i,j}$ from \eqref{eq: p-}, $\forall (i,j)$;\\
    Determine $\mathbb{I}^{(m)}_1$ from \eqref{eq: I1} and $\mathbb{I}^{(m)}_2$ from \eqref{eq: I2};\\
    
    \eIf{$\sum_{(i,j)\in\mathbb{I}^{(m)}_1} p^+_{i,j}  < \sum_{(i,j)\in\mathbb{I}^{(m)}_2} p^{-}_{i,j}$}{
    $\hat{m}_{i_1,j_1} \gets \hat{m}_{i_1,j_1}+2,\ \forall(i_1,j_1)\in\mathbb{I}^{(m)}_1$;\\
    $\hat{m}_{i_2,j_2} \gets \hat{m}_{i_2,j_2}-2, \forall(i_2,j_2)\in\mathbb{I}^{(m)}_2$;\\
    }{\textbf{Break};\\}
    }}
    }
    
    $v_{i^*} \leftarrow v_{i^*}+1$;\\
    $\hat{m}_{i,j} = \hat{m}_{i,j}^{\rm buffer}$, $\forall i,j$;\\
    \textbf{// Post processing}:\\
    Determine $\{(i_u,j_u)\}_{u=1}^{NB}$ by sorting $\{{\mu}_{i,j}^{(v_i)}\}_{i,j}$; \\
    $\mathcal{A} = \{1,...,NB\}$;\\
    \For{$t\in\{1,...,T\}$}{
    $u^* = \min(\mathcal{A})$;\\
    $m_t = \hat{m}_{(i_{u^*},j_{u^*})}$;\\
    $\mathcal{J}_t=\{u\in \mathcal{A}\ |\ \hat{m}_{(i_{u},j_{u})}=m_t\}$;\\
    $\text{Choose }\mathcal{K}_t \text{ as the set of }m_t\text{ the smallest indices in }\mathcal{J}_t$;\\
    $\bar{\mu}_t= \frac{1}{m_t}\sum_{(i,j)\in\mathcal{K}_t}\mu_{i,j}^{(v_{i})}$;\\
    $p_t =\hat{\epsilon}^{-1}(\bar{\mu}_t;m_t)$;\\
    $\mathcal{A} \gets \mathcal{A}-\mathcal{K}_t$;\\
    }
    $p_t\gets  p_t + \frac{1}{T}( P_{\rm tot}-\sum_{t=1}^T p_t)$;\\
    
    \end{algorithm}}}

Our second sub-problem primarily aims to minimize the overall distortion by optimizing the temporary modulation order matrix $\hat{\bf M}$ for a fixed ${\bf v}$.
When fixing the codebook assignment vector, the original objective function $\sum_{i=1}^{N}\hat{\mathcal{D}}_i(v_i)$ in \eqref{eq:Opt_1-2} becomes a constant. 
Therefore, we choose the total transmit power $\sum_{i=1}^N \sum_{j=1}^B \hat{p}_{i,j}^\star$ as the objective function of our second sub-problem as follows:
\begin{subequations}\label{eq:Opt_1-4}
\begin{align}
    \text{(P2)}~\min_{{\bf v}}~&\sum_{i=1}^N \sum_{j=1}^B \hat{p}_{i,j}^\star, \\ 
    \text{s.t.}~~& \hat{p}_{i,j}^\star  = \frac{\hat{\epsilon}^{-1}(\mu_{i,j}^{(v_i)} ; \hat{m}_{i,j})}{\hat{m}_{i,j}},  \\
    &\frac{1}{NB}\sum_{i=1}^N \sum_{j=1}^B \hat{m}_{i,j} \geq R, ~\hat{m}_{i,j} \in  \{2,4,...,m_{\rm max}\},\label{eq:Opt_1-4_C1}
\end{align}
\end{subequations}
This problem is still relevant to the original problem \eqref{eq:Opt_1-2} because reducing the total transmit power contributes to offer a room to reduce the overall distortion when solving the first sub-problem (P1). 
To solve the second sub-problem, we also employ a greedy-type swapping algorithm. In this algorithm, we minimize the total transmission power by making adjustments on temporary modulation orders while maintaining the constraint in \eqref{eq:Opt_1-4_C1}.
Basically, the algorithm is performed for the modulation order $m\in\{4,...,m_{\rm max}-2\}$. In each iteration, we identify index sets $\mathbb{I}^{(m)}_1,\mathbb{I}_2^{(m)}$ and examine whether the total power can be reduced by adjusting $\hat{m}_{i_1,j_1} \gets m +2$, $\forall(i_1,j_1)\in\mathbb{I}_1$ and $\hat{m}_{i_2,j_2} \gets m - 2$, $\forall(i_2,j_2)\in\mathbb{I}_2$. This adjustment leads to the increase in $\hat{p}_{i_1,j_1}^\star$ and the decrease in $\hat{p}_{i_2,j_2}^\star$. To maximize the power reduction, while minimizing the power increment, we identify index sets $\mathbb{I}^{(m)}_1$ and $\mathbb{I}^{(m)}_2$ as follows:
\begin{align}
\mathbb{I}^{(m)}_1 &= \{(i,j)\ |\ \hat{m}_{i,j}=m,\ p^+_{i,j}\leq{\sf sort}_{m+2}({\bf P}^+)\},\label{eq: I1}\\
\mathbb{I}^{(m)}_2 &= \{(i,j)\ |\ \hat{m}_{i,j}=m,\ p^-_{i,j}\geq{\sf sort}_{NB-(m-2)}({\bf P}^-)\},\label{eq: I2}
\end{align}
where $p^+_{i,j}$ and $p^-_{i,j}$ are defined as
\begin{align}
p^+_{i,j} =  \hat{\epsilon}^{-1}(\mu_{i,j}^{(v_i)} ; \hat{m}_{i,j}+2)-\hat{\epsilon}^{-1}(\mu_{i,j}^{(v_i)} ; \hat{m}_{i,j}),\label{eq: p+}\\
p^-_{i,j} = \hat{\epsilon}^{-1}(\mu_{i,j}^{(v_i)} ; \hat{m}_{i,j}) -\hat{\epsilon}^{-1}(\mu_{i,j}^{(v_i)} ; \hat{m}_{i,j} -2),\label{eq: p-}
\end{align}
and ${\bf P}^+$ and ${\bf P}^-$ are the power increment matrix and power decrement matrix, respectively, such that $[{\bf P}^+]_{i,j} =p^+_{i,j}$ and $[{\bf P}^-]_{i,j} =p^-_{i,j}$.
If the total amount of the power increment in $\mathbb{I}^{(m)}_1$ is smaller than the total amount of the power reduction in $\mathbb{I}^{(m)}_2$, the adjustment on the modulation orders $\hat{m}_{i_1,j_1},\ \forall(i_1,j_1)\in\mathbb{I}^{(m)}_1$ and $\hat{m}_{i_2,j_2},\ \forall(i_2,j_2)\in\mathbb{I}^{(m)}_2$ are made. This ensures that the rate constraint is maintained while minimizing the total transmit power.
This adjustment is performed iteratively until the amount of the power increment becomes larger than the amount of the power reduction.

\begin{figure}[t]
    \centering 
    {\epsfig{file=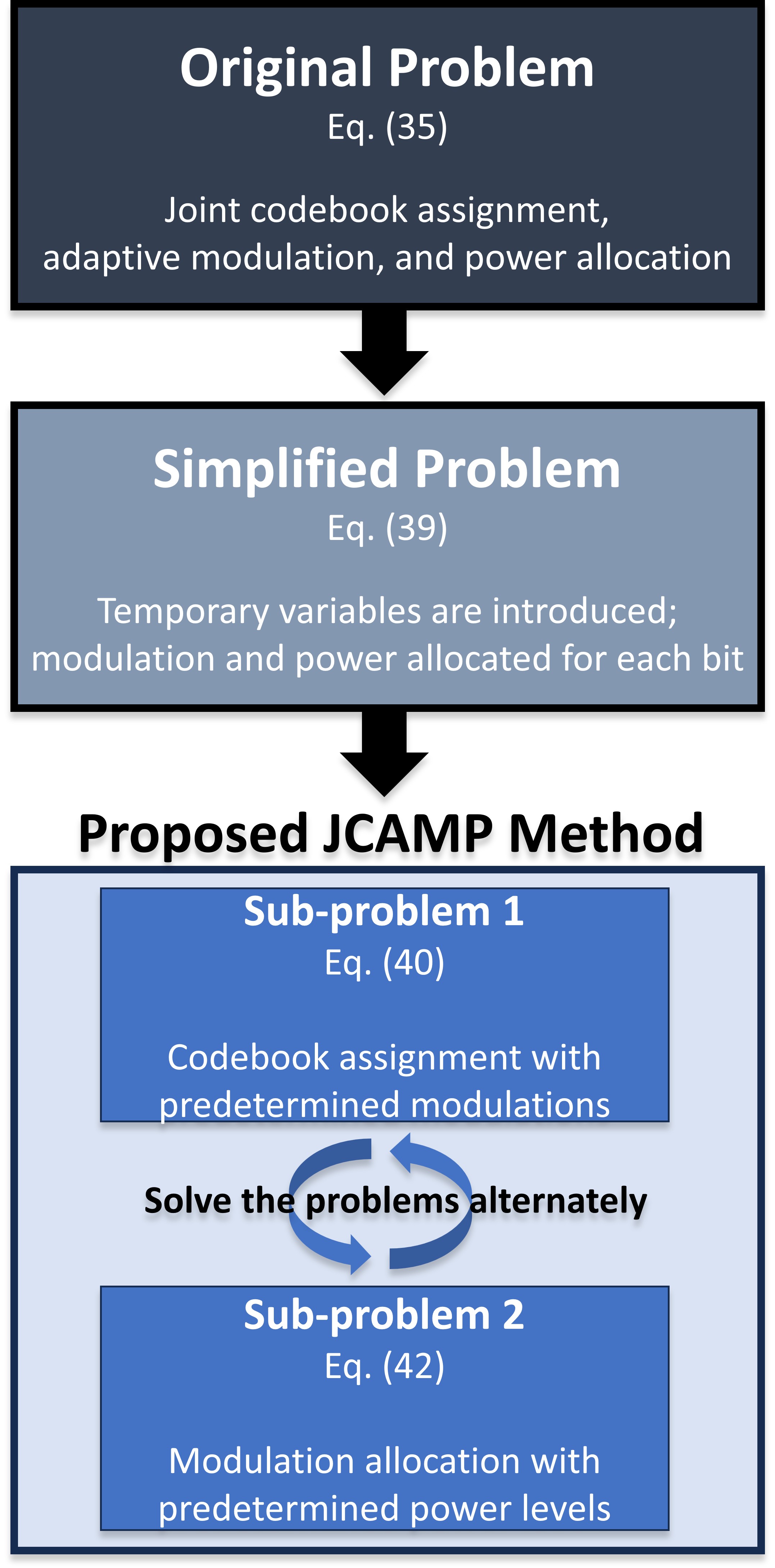, width=4.5cm}} 
    \caption{An illustration of the overall flow of the problem formulations for the proposed JCAMP method.}
    \label{fig:Overall Flow_JCAMP}\vspace{-3mm}
\end{figure}

In the proposed JCAMP method, the codebook assignment vector and temporary modulation order matrix are initialized as $v_i=V$ and $\hat{m}_{i,j}=\frac{NB}{T}$, $\forall i,j$. Then,  the codebook assignment vector ${\bf v}$ is updated by solving the sub-problem (P1) while keeping $\hat{\bf M}$ is fixed. Afterward,  the modulation order matrix $\hat{\bf M}$ is updated by solving the sub-problem (P2) while keeping ${\bf v}$ fixed. These steps are repeated until the temporary power constraint $\sum_{i=1}^N \sum_{j=1}^B \hat{\epsilon}^{-1}(\mu_{i,j}^{(v_i)} ; \hat{m}_{i,j}) \leq  P_{\rm tot}$ is satisfied. 
    
Once ${\bf v}$ and $\hat{\bf M}$ are determined, a post-processing step is required to determine the {\em actual} power allocation $\{ p_t\}_t$ and modulation orders $\{m_t\}_t$ for the $T$ transmitted symbols. 
First of all, the bit sequence is sorted in ascending order of the bit-flip probabilities $\{\mu_{i,j}^{(v_i)}\}_{i,j}$ determined from ${\bf v}$.
The index $(i_u,j_u)$ is then defined as the index associated with the $u$-th lowest bit-flip probability among $\{{\mu}_{i,j}^{(v_i)}\}_{i,j}$.
After sorting, bits with the same $\hat{m}_{i,j}$ values and similar $\mu^{(v_i)}_{i,j}$ values are grouped into $\{s_t\}_t$.
For each symbol $s_t$, let $\mathcal{A}$ be the set of ungrouped bit indices $\{(i_u,j_u)\}_u$. The modulation order $m_t$ for $s_t$ is then selected as $m_t = \hat{m}_{i_{u^*},j_{u^*}}$, where $u^*$ is the smallest index in $\mathcal{A}$.
Next, we identify the candidate set of bit indices with the same modulation order:
\begin{gather}
\mathcal{J}_t=\{u\in \mathcal{A}\ |\ \hat{m}_{i_{u},j_{u}}=m_t\},
\end{gather}
and define $\mathcal{K}_t$ as the $m_t$ smallest indices in $\mathcal{J}_t$.
This index set $\mathcal{K}_t$ is used for symbol mapping at the transmitter and for symbol de-mapping at the receiver\footnote{During this process, no exchange of bit grouping information between the transmitter and receiver is necessary, as both ends operate the same algorithm under shared communication constraints, yielding identical grouping results.}\footnote{ Moreover, since both ends are aware of the bit grouping results, it is also feasible to implement an additional symbol permutation algorithm to minimize the hardware burden, which is caused by frequent modulation order and power switching. The transmitter can permute the symbols so that those with similar modulation orders and power levels are adjacent, and the receiver can perform the corresponding de-permutation after detecting the symbols.}.
Finally, the power allocated to $s_t$ is computed as $p_t = \hat{\epsilon}^{-1}(\bar{\mu}_t;m_t)$, where $\bar{\mu}_t$ is the average bit-flip probability for the bits in $\mathcal{I}_t$, computed as
\begin{align}\label{eq: bar_mu}
    \bar{\mu}_t= \frac{1}{m_t}\sum_{(i,j)\in\mathcal{K}_t}\mu_{i,j}^{(v_{i})}.
\end{align}
This process is repeated until all the $NB$ bits are assigned appropriate modulation orders and power levels.
If the total power constraint is not satisfied due to discrepancies between temporary and actual power, the power allocation is further adjusted as
\begin{align}\label{eq: post_processing}
p_t\gets  p_t + \frac{1}{T}( P_{\rm tot}-\sum_{t=1}^T p_t).
\end{align}
The overall procedure of the proposed JCAMP method, including the post-processing step, is summarized in Algorithm~\ref{alg:JCAMP}.  The overall flow of the problem formulations for the proposed JCAMP method is illustrated in Fig. \ref{fig:Overall Flow_JCAMP}.

In practice, the iterative optimization process in the JCAMP method can incur significant computational cost, especially in environments constrained by processing power or low-latency demands. To mitigate this challenge, a {\em lookup table}-based strategy can be employed, where the instantaneous SNR is discretized into fixed intervals and the corresponding optimal power and modulation allocations are pre-calculated offline. This approach removes the need for real-time optimization during communication, allowing for efficient execution with negligible complexity.

\subsection{Joint Codebook Assignment and Power Allocation (JCAP) Method}\label{Sec:JCAP}
The proposed JCAMP method in Sec.~\ref{Sec:JCAMP} jointly optimizes codebook assignment, modulation order, and power allocation. However, adjusting the modulation order across the bit sequence may increase both implementation and computation complexities. 
To reduce these complexities, we also present a joint codebook assignment and power allocation (JCAP) method that solve sthe optimization problem in \eqref{eq:Opt_0} under fixed modulation. 
In this method, we consider a fixed modulation order $m_t = R = \frac{NB}{T}$, implying that the constraint in \eqref{eq:Opt_0_C3} is satisfied.
Then, the original problem \eqref{eq:Opt_0} simplifies to  
\begin{subequations}\label{eq:Opt_2}
\begin{align}
    \min_{{\bf v},{\bf p}}~&  \sum_{i=1}^{N}\hat{\mathcal{D}}_i(v_i), \\
    \text{s.t.}~&\hat{\epsilon}(p_t,R) = \frac{1}{R}\sum_{u=(t-1)R+1}^{t R} {\sf sort}_u(\tilde{\bm \mu}({\bf v})),~\forall t, 
    \label{eq:Opt_2_C1} \\
    & \sum_{t=1}^T p_t \leq  P_{\rm tot},~p_t\geq 0. 
\end{align}
\end{subequations}
This problem aims at jointly optimizing the codebook assignment vector ${\bf v}$ and power allocation vector ${\bf p}$ to minimize the overall distortion. 

Solving \eqref{eq:Opt_2} is still challenging due to the entanglement between ${\bf v}$ and ${\bf p}$. To circumvent this challenge, we temporarily assume that the BER-matching constraint is given in a per-bit basis as follows:
\begin{align}\label{eq:BER_match_temp2}
    \hat{\epsilon}(\hat{p}_{i,j}R;R) = \mu_{i,j}^{(v_i)},
\end{align}
where $\hat{p}_{i,j}$ is referred to as a temporary power.
As discussed in Sec.~\ref{Sec:JCAMP}, the temporary power can be efficiently determined by solving the equation \eqref{eq:BER_match_temp2} for $\hat{p}_{i,j}$ using numerical methods. 
Then the corresponding solution is obtained as $\hat{p}_{i,j}^\star = \hat{\epsilon}^{-1}(\mu_{i,j}^{(v_i)};R)/R$.
Based on the temporary power allocation, our problem \eqref{eq:Opt_2} reduces to a codebook assignment problem:
\begin{subequations}\label{eq:Opt_2-2}
\begin{align}
    \min_{{\bf v}} ~&\sum_{i=1}^{N}\hat{\mathcal{D}}_i(v_i), \\ 
    \text{s.t.}~&\sum_{i=1}^N \sum_{j=1}^B \hat{p}_{i,j}^\star  \leq  P_{\rm tot},~ \hat{p}_{i,j}^\star  = \frac{1}{R}\hat{\epsilon}^{-1}(\mu_{i,j}^{(v_i)} ;R). \label{eq:Opt_2-2_C1}
\end{align}
\end{subequations}
To solve the above problem in an efficient manner, we devise a greedy allocation algorithm, as done in Sec.~\ref{Sec:JCAMP}. In this algorithm, we initialize the codebook assignment vector by choosing the highest codebook index (i.e., $v_i=V,\ \forall i$). Then, during iterations, we gradually reduce certain $v_i$ values step-by-step to meet the power constraint. In particular, at each iteration, we select the index $i^*$ that maximizes the distortion reduction relative to power increase:  
\begin{align}\label{eq:JCAP_criterion}
    i^*=\argmax_{i:v_i>1} \frac{\hat{\mathcal{D}}_i(v_i) -\hat{\mathcal{D}}_i(v_i-1)}{\sum_{j=1}^B \big\{ \hat{\epsilon}^{-1}(\mu_{i,j}^{(v_i-1)} ; R) - \hat{\epsilon}^{-1}(\mu_{i,j}^{(v_i)} ; R)\big\}},
\end{align}

{\small \RestyleAlgo{ruled}
    \SetKwComment{Comment}{/* }{ */}
    \begin{algorithm}[t]\label{alg:JCAP}
    \caption{Proposed JCAP method}
    $v_i = V$, $\forall i$, $R=\frac{NB}{T}$;\\
    \While{$\sum_{i,j} \hat{\epsilon}^{-1}(\mu_{i,j}^{(v_i)};R) \leq P_{\rm tot}R$}{
        Choose $i^*$ from \eqref{eq:JCAP_criterion};\\
        $v_{i^*} \leftarrow v_{i^*}-1$;\\
    }
    $v_{i^*} \leftarrow v_{i^*}+1$;\\
    {\bf // Post processing}: \\
    Determine $\{(i_u,j_u)\}_{u=1}^{NB}$ by sorting $\{{\mu}_{i,j}^{(v_i)}\}_{i,j}$; \\
    \For{$t\in\{1,\ldots,T\}$}{
    $\mathcal{K}_t=\{(i_{(t-1)R+1},j_{(t-1)R+1}),\ldots,(i_{tR},j_{tR})\}$;\\
    $\bar{\mu}_t= \frac{1}{R}\sum_{(i,j)\in\mathcal{K}_t}\mu_{i,j}^{(v_{i})}$;\\
    $p_t =\hat{\epsilon}^{-1}(\bar{\mu}_t;R)$;\\
    }
    $p_t\gets  p_t + \frac{1}{T}( P_{\rm tot}-\sum_{t=1}^T p_t)$;\\
    \end{algorithm}}

Once the codebook assignment vector ${\bf v}$ is determined using the greedy algorithm, the actual power allocation $p_t$ is obtained by the same post-processing as in the previous subsection with a fixed modulation order. The bit sequence is sorted in ascending order of the bit-flip probabilities $\{\mu^{(v_i)}_{i,j} \}_{i,j}$ determined from ${\bf v}$. The index $(i_u,j_u)$ is then defined as the index associated with the $u$-th lowest bit-flip probability among $\{{\mu}_{i,j}^{(v_i)}\}_{i,j}$. Following the sorted order, we define $\mathcal{K}_t=\{(i_{(t-1)R+1},j_{(t-1)R+1}),\ldots,(i_{tR},j_{tR})\}$ as the set of bit indices grouped into symbol $s_t$. Afterwards, for the average bit-flip probability $\bar{\mu}_t$ defined in \eqref{eq: bar_mu}, we compute $p_t = \hat{\epsilon}^{-1}(\bar{\mu}_t;m_t)$. To resolve the discrepancy between the temporarily computed power under the constraint and the actual total $\sum^T_{t=1}p_t$, the power adjustment in \eqref{eq: post_processing} is applied. The overall procedure of the proposed JCAP method is summarized in Algorithm \ref{alg:JCAP}. As mentioned in the previous subsection, the proposed JCAP method, like the JCAMP method, can also adopt a lookup table-based approach to alleviate the associated computational burden.

\section{Simulation Results and Analysis}\label{Sec:Simul}
In this section, we demonstrate the superiority of the proposed ESC-MVQ framework with simulation results. In our simulations, we focus on image reconstruction tasks where the task-dependent loss is set as
$d_0(\mathbf{x},\mathbf{\hat x})= \Vert \mathbf{x} - \mathbf{\hat x} \Vert^2$. For the quantization loss, we adopt the loss structure in \cite{VQ-VAE} given by $d_1(\mathbf{z},\hat{\bf z}_q) = \left\Vert \text{sg}(\mathbf z)-\hat{\bf z}_q\right\Vert^2 + \beta \left\Vert\mathbf z-\text{sg}(\hat{\bf z}_q)\right\Vert^2,$ where $\beta=0.25$ is a hyperparameter. The quantization loss aims to minimize the difference between the semantic feature $\mathbf{z}$ and its quantization $\hat{\bf z}_q$. For the performance measure, we adopt the peak signal-to-noise ratio (PSNR) defined as
\begin{align}
{\rm PSNR} = 10~{\rm log}_{10}\left( \frac{{\rm MAX}^2}{{\rm MSE}^2} \right)\ ({\rm dB}),
\end{align}
where MAX denotes the maximum possible pixel value of the image (e.g., 255 for 8-bit image), and MSE represents the mean-squared-error between the input and reconstructed images.
The normalized signal-to-noise ratio (SNR) for transmitting the entire bit sequence is defined as
\begin{align}\label{eq: SNR}
{\sf SNR} = 10~{\rm log}_{10}\left( \frac{P_{\rm tot}\mathbb{E}[\gamma]}{NB} \right)\ ({\rm dB}).
\end{align}

    

We consider the CIFAR-10, CIFAR-100, and STL-10 datasets.
In the training process, the batch size and number of epochs are set as 64 and 128, respectively. Also, the Adam optimizer is employed with a learning rate of $10^{-3}$ to train the semantic encoder-decoder. For the encoder-decoder networks, the network structure in \cite{VQ-VAE} is adopted. The compression ratio is defined as 
\begin{align}\label{eq: compression ratio}
\rho=\frac{\text{(transmission bits)}}{C\times H \times W\times 8},
\end{align}
where $C$, $H$, and $W$ denote the channel, height, and width of the input image, respectively. In our simulations, the CIFAR-10 and CIFAR-100 datasets have dimensions $(C,H,W)=(3,32,32)$, while the STL-10 dataset has dimensions $(3,96,96)$. In our simulations, $\rho=3/64$ was used, which corresponds to a transmission overhead of 1152 bits for the CIFAR-10 and CIFAR-100 datasets, and 10368 bits for the STL-10 dataset.
Additionally, the rate constraint $R$ is set as 4.
For performance comparison, we consider the following communication frameworks:

\begin{figure*}[t]
    \centering
    \subfigure[CIFAR-10]
    {\epsfig{file=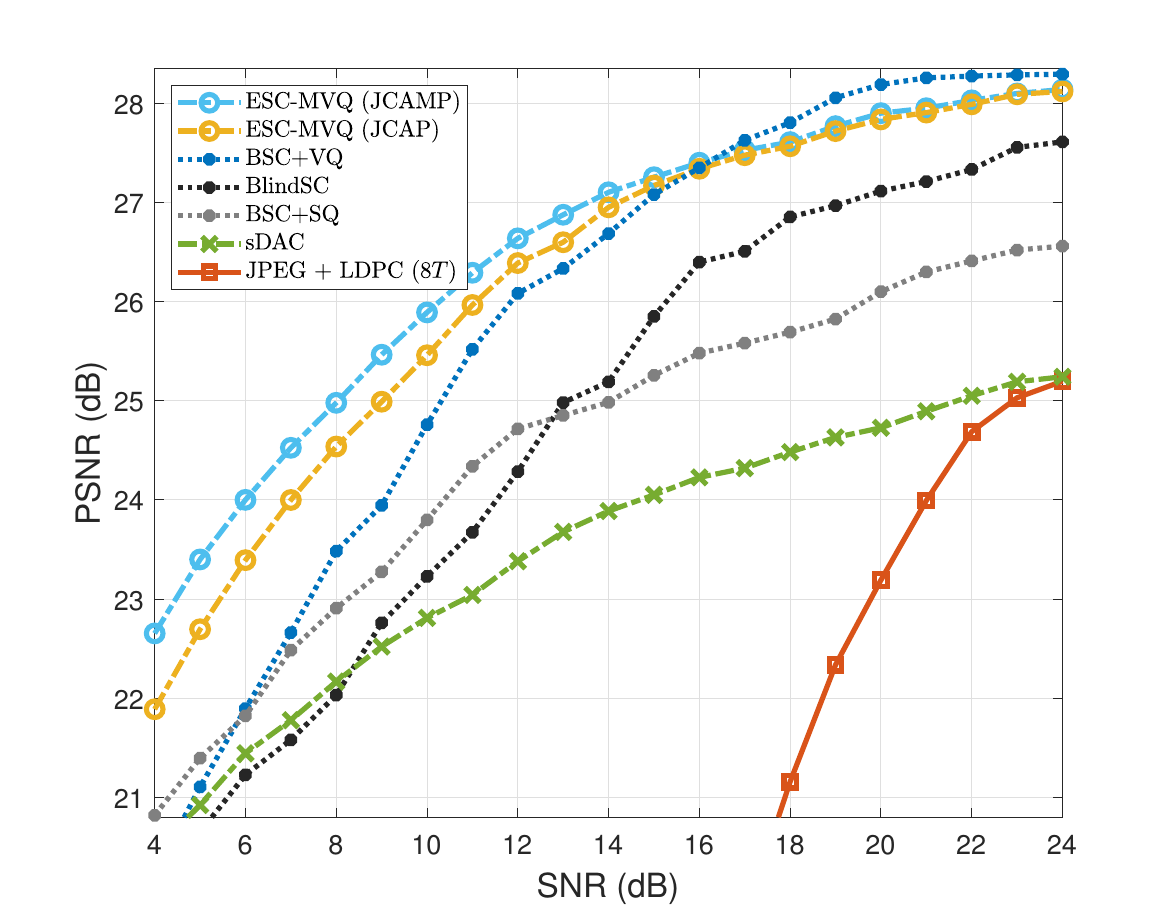,width=6cm}}
    \subfigure[CIFAR-100]
    {\epsfig{file=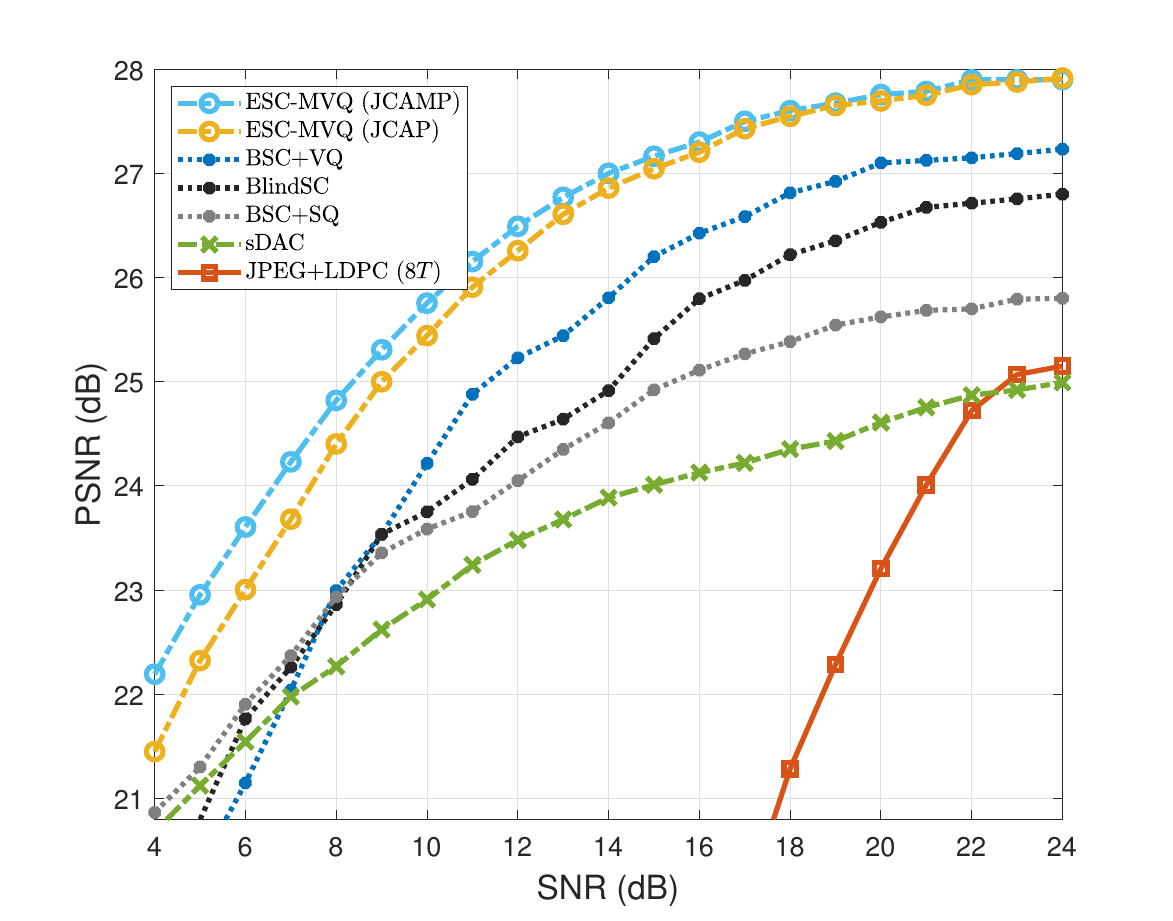,width=6cm}}
    \subfigure[STL-10]
    {\epsfig{file=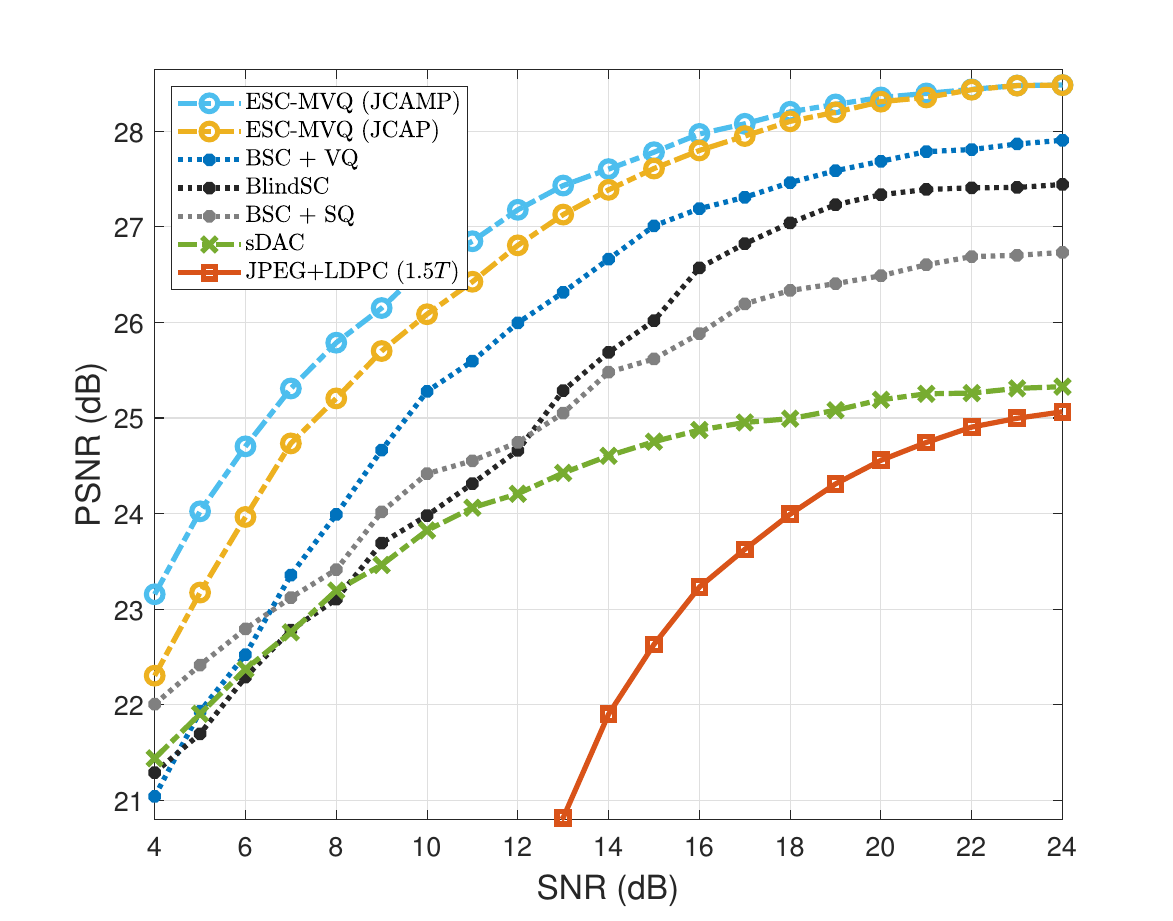,width=6cm}}
    \caption{PSNR performances of various communication frameworks across different SNRs in a Rayleigh fading channel.}\vspace{-3mm}
    
    \label{fig:PSNR_Rayleigh}
\end{figure*}

\begin{itemize}
    \item {\bf ESC-MVQ (JCAMP, JCAP)}: This is the proposed ESC-MVQ framework combined with the JCAMP method in Sec. \ref{Sec:JCAMP} or the JCAP method in Sec. \ref{Sec:JCAP}. For the multi-codebook training strategy in Sec. \ref{Sec:Multi-Codebook}, the number of codebooks is set as $V=5$ unless specified otherwise.  Other parameters are set as $D=4$, $B=9$, $\eta=0.8$, $[\mu^{(1)}_{\rm min},...,\mu^{(5)}_{\rm min}]=[0.0005,0.001,0.0045,0.02,0.05]$, and $\lambda^{(v)}=\frac{1}{8}\times2^{v-1},\ \forall v\in\{1,...,5\}$. The parameter $\tau$ in \eqref{eq: gumbel_softmax_multi} is initially set as 0.5 and then annealed as $\tau \times e^{-0.003}$ for every 100 training iterations. The maximum modulation level is set to $m_{\rm max}=6$.
    The bit-flip probabilities are optimized using the Adam optimizer with a learning rate of $10^{-2}$. This framework utilizes only a single encoder-decoder pair for all simulations.
    When the channel gain $|h|^2$ is extremely low, assigning the $V$-th codebook to all sub-vectors may still exceed the power constraint. In such cases, JCAMP and JCAP normalize the total power by scaling it down to satisfy the constraint.

    \item {\bf BlindSC}: This is the conventional digital semantic communication framework developed in \cite{BlindJSCC}. In this framework, the encoder-decoder pairs are trained for the parallel BSCs, where the bit-flip probabilities of the BSCs are jointly optimized. Unlike ESC-MVQ, this framework employs a predefined uniform SQ module. In our simulations, the number of quantization bits is set as 3, which maximizes performance while maintaining the same compression ratio as in ESC-MVQ. This framework employs the MSE between $\bar{\bm \mu}$ and 0.5 for $\mathcal{R}(\bar{\bm \mu})$. The bit-flip probabilities are optimized using the Adam optimizer with a learning rate of $10^{-2}$. To meet different power constraints under various channel conditions, this framework requires $K$ different encoder-decoder pairs. 

    \item {\bf BSC+VQ}: This is a simple variation of the proposed ESC-MVQ framework. In this framework, the encoder-decoder pairs are trained for the parallel BSCs with a fixed (and non-trainable) bit-flip probability $\mu_{i,j}=\mu_{\rm const}\in\{0.001,0.0045,0.02,0.05,0.1\}$, $\forall i,j$. To meet different power constraints under various channel conditions, this framework requires $K$ different encoder-decoder pairs.
    
    \item {\bf BSC+SQ}: This is the conventional digital semantic communication framework developed in \cite{NECST}. In this framework, the encoder-decoder pairs are trained for the parallel BSCs with a fixed (and non-trainable) bit-flip probability. Unlike ESC-MVQ, this framework employs a predefined uniform SQ module. In our simulations, the number of quantization bits is set as 3, which maximizes performance while maintaining the same compression ratio as in ESC-MVQ. To meet {\em every} power constraints, this framework requires $K$ different encoder-decoder pairs {\em seperately} trained for each power constraint.

    \item {\bf sDAC}: This is a digital semantic communication framework proposed in \cite{sDAC}. The framework implements a digital analog converter tailored for semantic communication. It employs a learnable VQ module to convert semantic features into bit sequences and models the communication error using the BSC model. To enhance robustness across diverse channel conditions, the bit-flip probability of the BSC model is randomly sampled from $\mathcal{U}[0,0.1]$ at each training iteration. Reflecting this sampling strategy, the entire model is trained using a separate average semantic error (ASE) metric defined in \cite{sDAC}.

    \item {\bf JPEG+LDPC}: This is a traditional separate source-channel coding framework. This framework employs the joint photographic experts group (JPEG) for source coding and the 3/4-rate low-density parity-check (LDPC) code for channel coding. To ensure a fair performance comparison with existing frameworks, the bit and symbol lengths are set to be 8 times longer for the CIFAR-10 and CIFAR-100 datasets, and 1.5 times longer for the STL-10 dataset, relative to other methods. To accommodate this extended symbol length, the modulation order for each image is adaptively adjusted within the range of 2 to 6. 

\end{itemize}

    

    

First, we evaluate the superiority of the proposed ESC-MVQ framework compared to existing communication frameworks.
To this end, in Fig.~\ref{fig:PSNR_Rayleigh}, we compare the PSNR performance of various communication frameworks across different SNRs in a Rayleigh fading channel. To cover the entire SNR range, BSC+VQ and BlindSC employ $K=3$ separately trained encoder-decoder pairs, while BSC+SQ requires $K=21$ pairs, one for each point in the figure. In contrast, the proposed ESC-MVQ and sDAC utilize only a single encoder-decoder pair. To highlight the memory efficiency of ESC-MVQ, Table~I presents the number of parameters required by each scheme. As shown in this table, the proposed ESC-MVQ requires the fewest parameters, owing to the use of a single shared encoder-decoder pair.

\begin{figure}[t]
    \centering
    {\epsfig{file=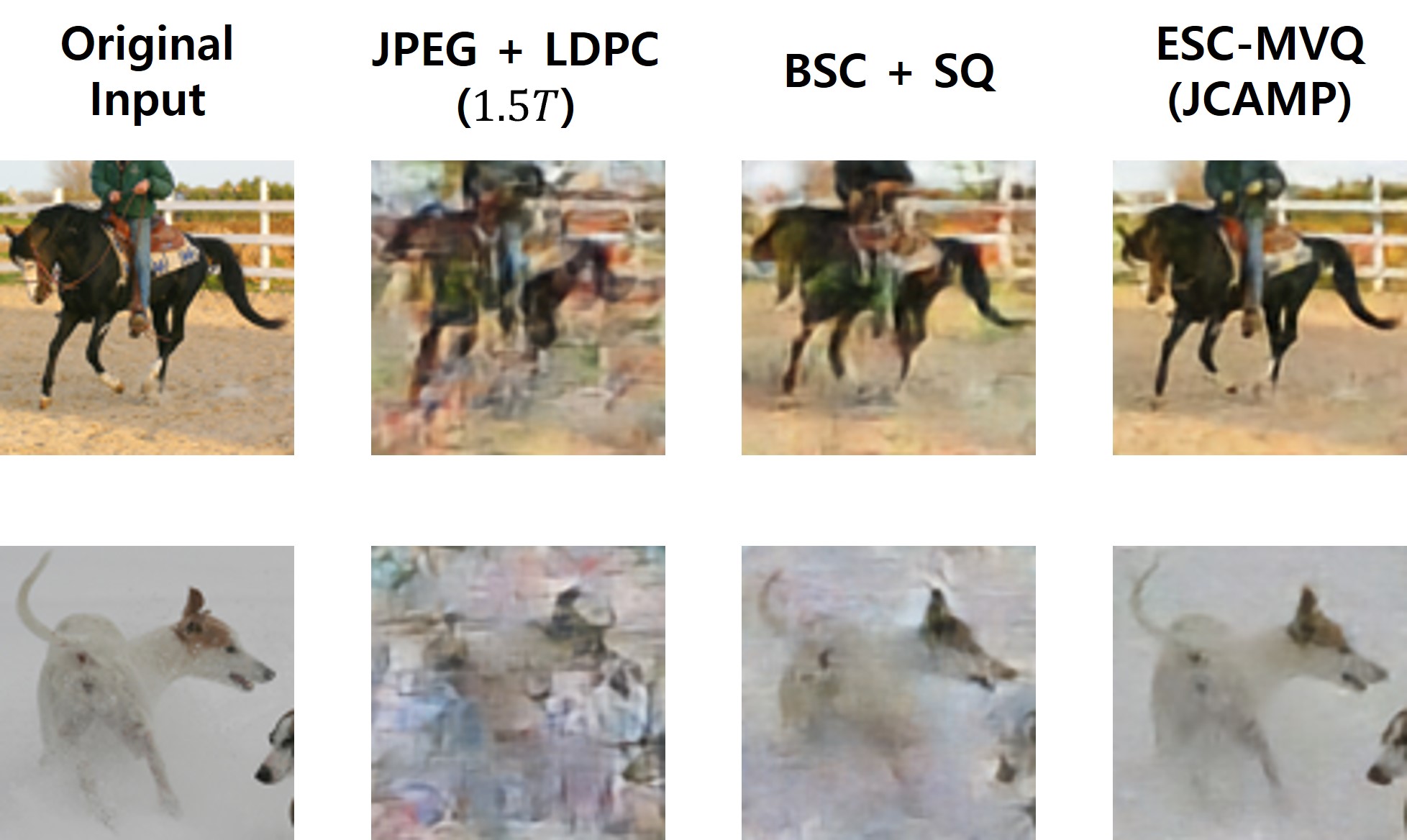,width=6cm}}
    \caption{Comparison of image reconstruction results across various communication frameworks on the STL-10 dataset under a Rayleigh fading channel with SNR = 10 dB.}\vspace{-3mm}
    \label{fig: image_visualization}
\end{figure}

As shown in Fig. \ref{fig:PSNR_Rayleigh}, ESC-MVQ, when integrated with the proposed JCAMP or JCAP method, significantly outperforms the baseline methods. These results not only validates that the proposed multi-codebook training strategy enables a single model to effectively operate over a wide SNR range, but also demonstrates that the JCAMP and JCAP methods successfully optimize the communication strategy in accordance with the trained encoder, VQ codebooks, and decoder. the JCAMP method achieves a notable performance gain, particularly in the low-SNR regime, by jointly adapting the modulation order alongside codebook assignment and power allocation. JPEG+LDPC exhibits inferior performance, while requiring a longer transmission delay compared to digital semantic communication frameworks. This highlights the significant advantage of joint source-channel coding in the digital semantic communication. Fig. \ref{fig: image_visualization} visualizes the image reconstruction performance of various comparison methods. It clearly shows that the proposed ESC-MVQ achieves significantly superior reconstruction performance compared to other communication techniques.

\begin{figure}[t]
    \centering
    {\epsfig{file=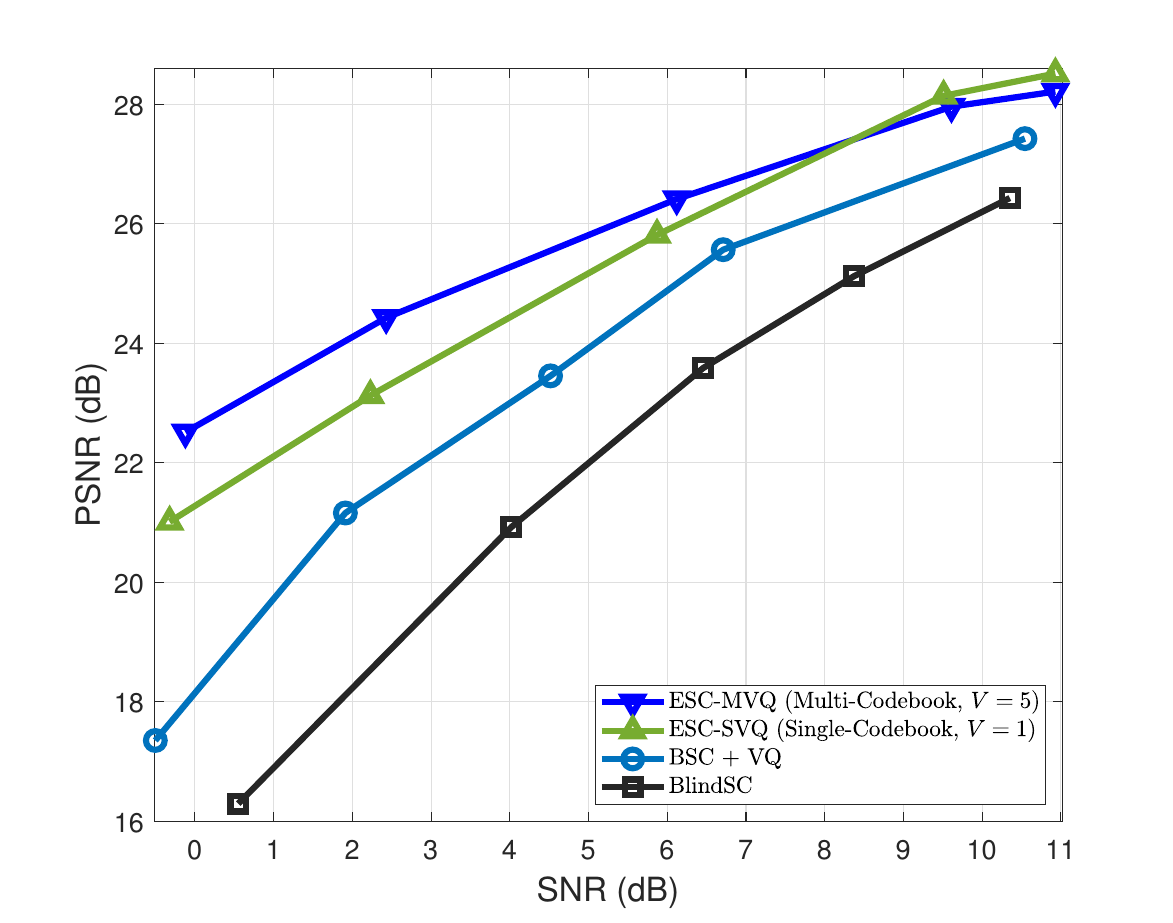,width=6cm}}
    \caption{PSNR performances of various training strategies across five different SNR levels in an AWGN channel using the STL-10 dataset.}\vspace{-3mm}
    \label{fig:PSNR_AWGN}
\end{figure}

Now, we demonstrate the superiority of the proposed multi-codebook training strategy compared to other training strategies. To this end, in Fig. \ref{fig:PSNR_AWGN}, we compare the PSNR performance of various training strategies across five different SNR levels in an AWGN channel using the STL-10 dataset. 
ESC-SVQ represents the single-codebook variant of the proposed strategy. To isolate and evaluate the impact of the training strategy, all considered strategies employ a fixed 16-QAM modulation scheme during inference, with appropriate power allocation to satisfy the BER-matching condition. It should be noted that, although all considered strategies employ parallel BSCs during end-to-end training, only ESC-MVQ leverages a single encoder-decoder pair. Unlike ESC-MVQ, other strategies rely on five separate encoder-decoder pairs, each specifically trained for a different SNR level.

Fig. \ref{fig:PSNR_AWGN} clearly shows that ESC-MVQ significantly outperforms other strategies. In particular, ESC-MVQ achieves comparable performance to its single-codebook counterpart in the high-SNR regime, while even surpassing it in the low-SNR regime. Considering that ESC-SVQ requires training five distinct encoder-decoder pairs, these results highlight the efficiency and superiority of our multi-codebook training strategy with $V=5$. Moreover, the performance gap between ESC-MVQ and BSC+VQ illustrates the benefits of learning the bit-flip probability in an end-to-end fashion, rather than relying on predefined values. This result emphasizes the importance of our strategy for jointly optimizing the bit-flip probability during training. Finally, the observed gap between ESC-MVQ and BlindSC highlights the advantage of incorporating a learnable VQ module for feature quantization, which enhances both memory efficiency and reconstruction quality.

\begin{table}[t]\label{table:parameters}
\caption{A comparison of the number of parameters used by the proposed scheme and the baseline methods.}
\centering
\begin{tabular}{c|cc}
\hline
\multirow{2}{*}{Scheme} & \multicolumn{2}{c}{\# of Parameters} \\ \cline{2-3} 
 & \multicolumn{1}{c|}{CIFAR-10/CIFAR-100} & STL-10 \\ \hline
\textbf{ESC-MVQ} & \multicolumn{1}{c|}{\textbf{524,427}} & \textbf{570,507} \\
Blind SC & \multicolumn{1}{c|}{1,521,051} & 1,548,699 \\
BSC + VQ & \multicolumn{1}{c|}{1,531,425} & 1,531,425 \\
BSC + SQ & \multicolumn{1}{c|}{10,623,165} & 10,623,165 \\
sDAC & \multicolumn{1}{c|}{647,413} & 647,413 \\ \hline
\end{tabular}
\end{table}

\begin{figure}[t]
    \centering 
    {\epsfig{file=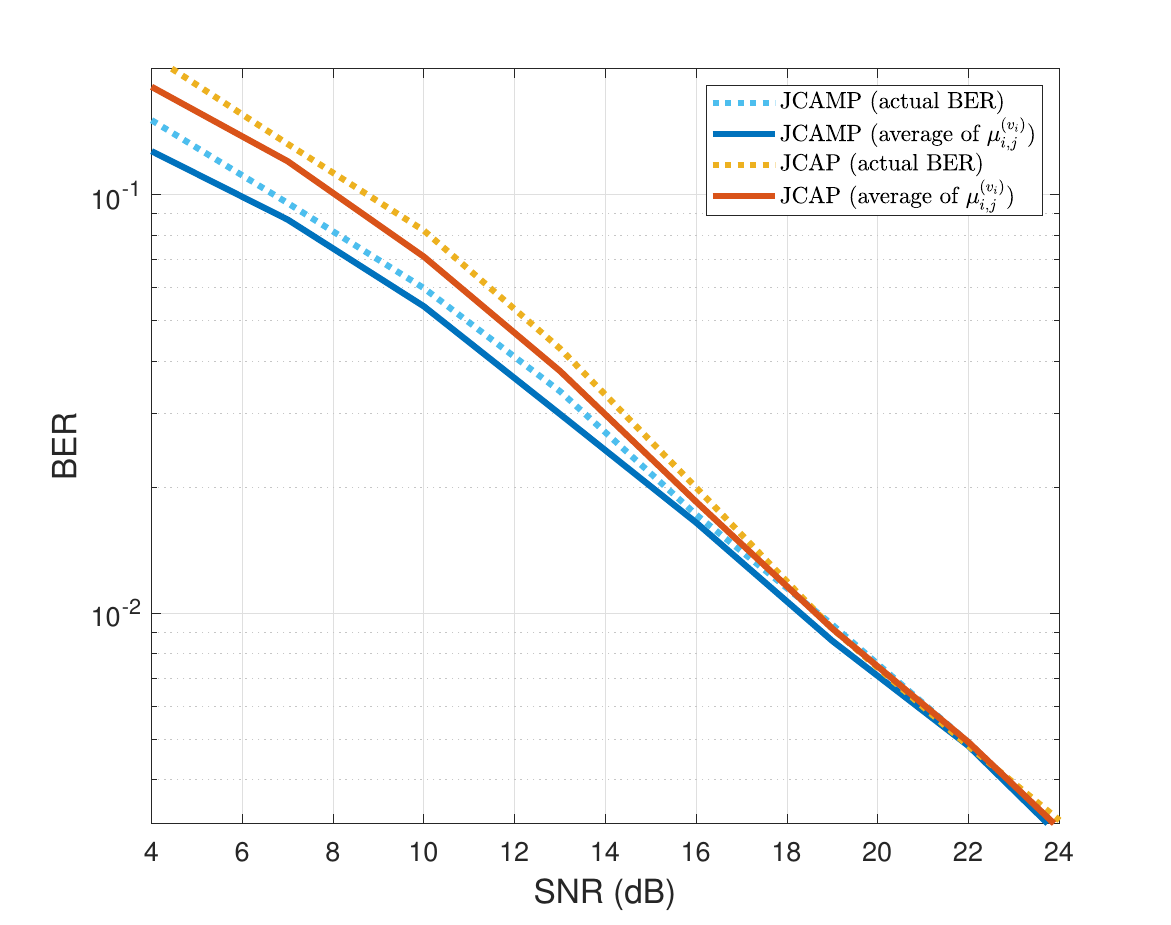, width=6cm}} 
    \caption{Comparison of average $\mu_{i,j}^{(v_i)}$
  values and actual measured BERs for the JCAMP and JCAP methods across various SNRs in a Rayleigh fading channel with the CIFAR-100 dataset.}
    \label{fig:BER1}\vspace{-3mm}
\end{figure}


\begin{figure}[t]
    \centering 
    {\epsfig{file=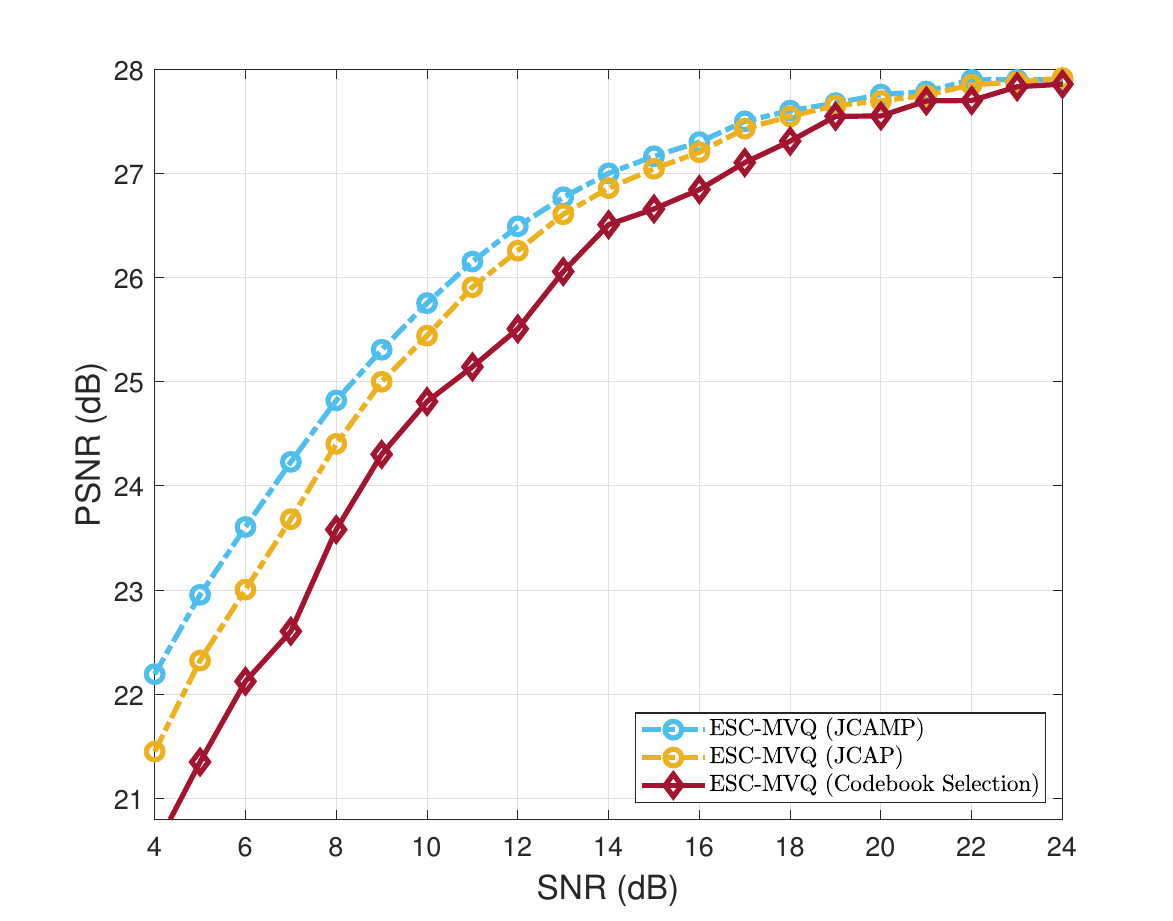, width=6cm}} 
    \caption{Performance gain of the proposed communication strategy over the naive usage of a trained ESC-MVQ model in a Rayleigh fading channel using the CIFAR-100 dataset.}
    \label{fig:Codebook Selection}\vspace{-3mm}
\end{figure}

\begin{figure}[t]
    \centering
    {\epsfig{file=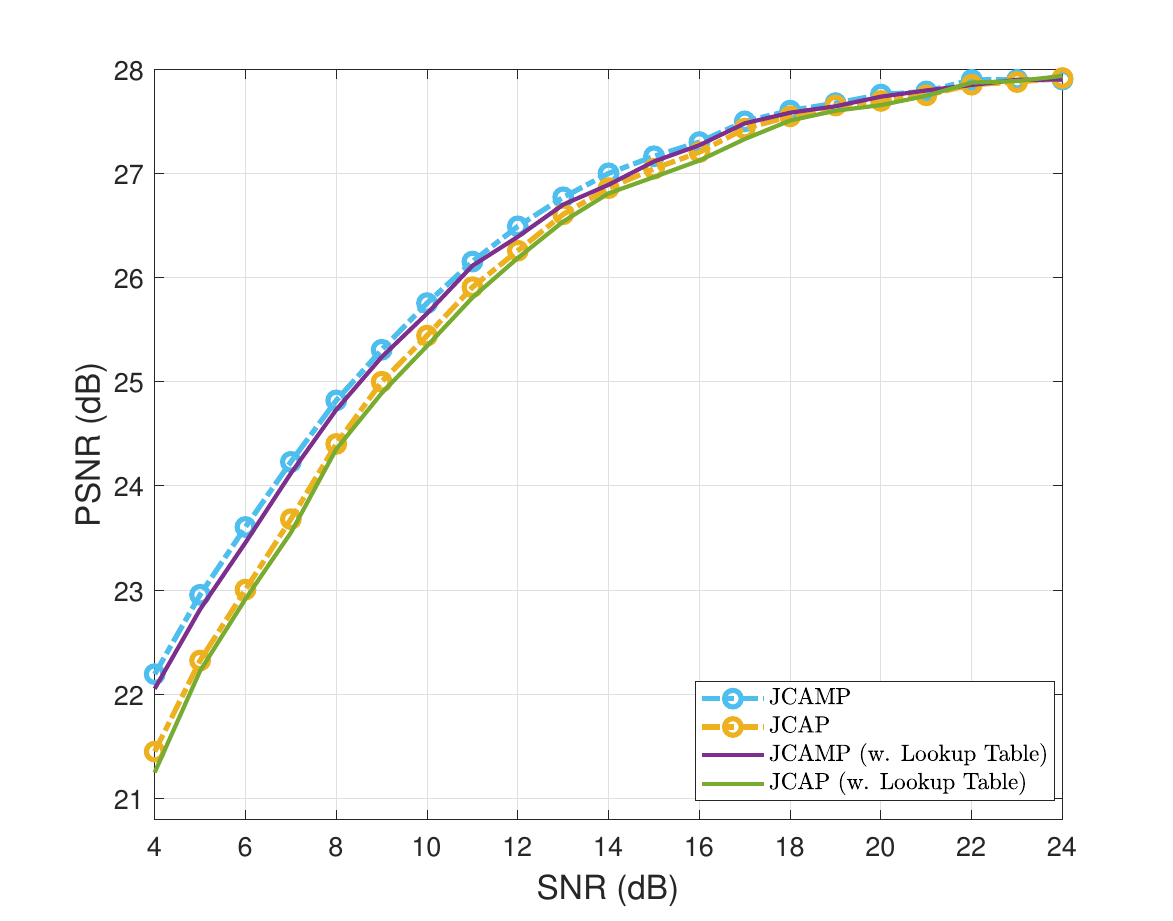,width=6cm}}
    \caption{PSNR performances of the JCAMP and JCAP methods with and without a lookup table for the CIFAR-100 dataset.}\vspace{-3mm}
    \label{fig: LookUpTable}
\end{figure}

\begin{figure}[t]
    \centering
    {\epsfig{file=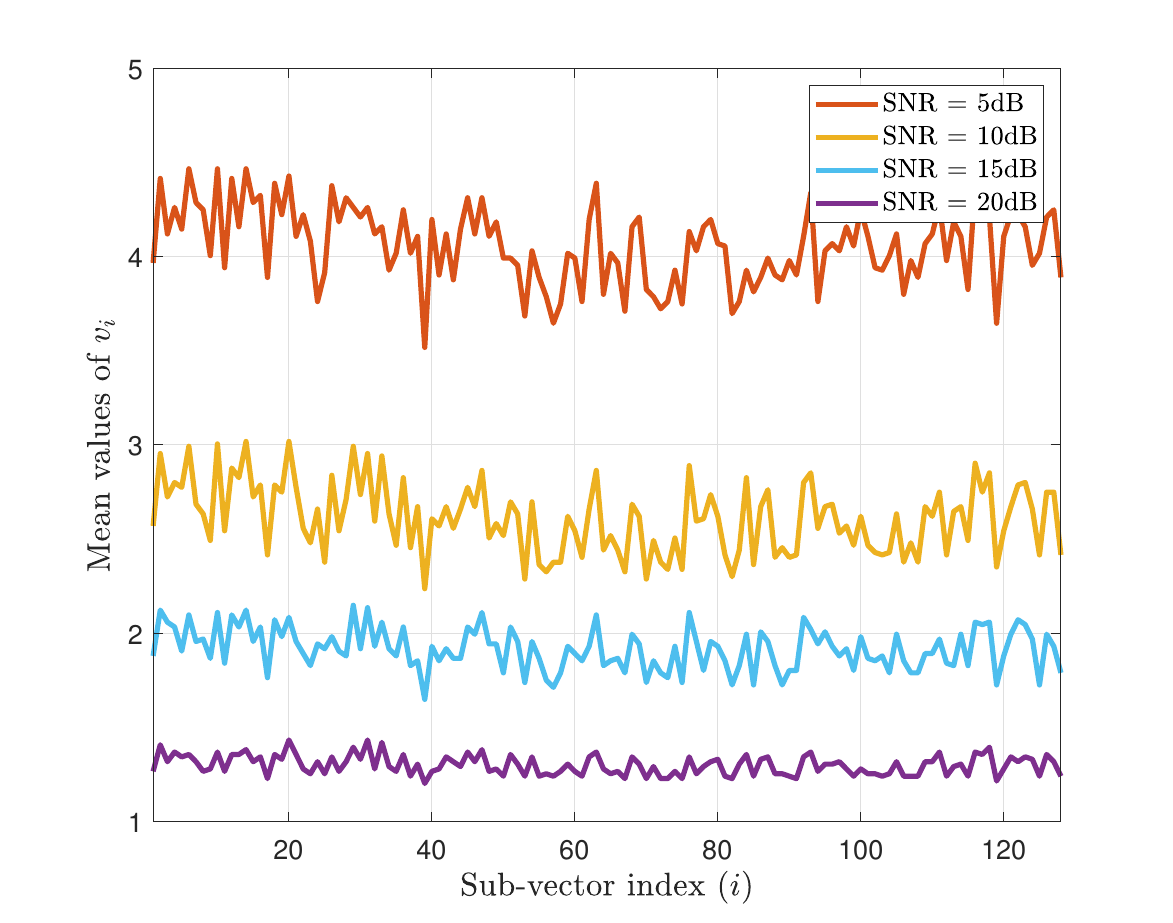,width=6cm}}
    \caption{Mean of the assigned codebook index for each sub-vector across different SNRs using the JCAMP method over a Rayleigh fading channel with the CIFAR-100 dataset.}\vspace{-3mm}
    \label{fig: codebook_usage}
\end{figure}

To validate the effectiveness of the proposed BER-matching condition, we compare the average values of $\mu^{(v_i)}_{i,j}$ with the actual measured BERs under the same experimental setup as in Fig. \ref{fig:PSNR_Rayleigh}. Fig. \ref{fig:BER1} presents the results for the JCAMP and JCAP methods and corresponds to the experimental setup used in Fig. \ref{fig:PSNR_Rayleigh}. Fig. \ref{fig:BER1} shows that the actual BERs closely align with the average bit-flip probabilities learned during training, demonstrating consistency between the training and inference environments. These results confirm that the proposed BER-matching condition is effectively established and maintained in practice.

To demonstrate the validity of the optimal communication strategy in Sec.~\ref{Sec:infer}, we present the performance gain of the proposed communication strategy over the naive usage of a trained ESC-MVQ model in a Rayleigh fading channel using the CIFAR-100 dataset. The baseline method in Fig. \ref{fig:Codebook Selection}, referred to as {\em Codebook Selection}, assumes a fixed modulation order $m_t=R$, and selects one of the codebook $v\in\{1,...,V\}$ for all sub-vectors based on the channel condition and power constraint. Power is then allocated according to the resulting bit-flip probabilities. Unlike our proposed strategy, which assigns different codebooks to each sub-vector and allocates modulation and power accordingly, this method corresponds to simply selecting a pre-trained model based on the channel condition—the approach used in BlindSC \cite{BlindJSCC}. Fig. \ref{fig:Codebook Selection} shows that our proposed JCAMP and JCAP methods outperform the {\em Codebook Selection} baseline. It demonstrates the efficacy of the proposed communication methods in Sec. IV.

Fig. \ref{fig: LookUpTable} demonstrates the validity of the lookup table-based JCAMP and JCAP methods, discussed in Secs. IV-B and IV-C, respectively. To construct the lookup table, we first quantize the {\em instantaneous} SNR, defined as $10\ {\rm log}_{10} \left(\frac{P_{\rm tot}\gamma}{NB}\right)\ ({\rm dB})$ , into discrete levels. For each level, we precompute the optimal power and modulation settings offline. All the quantized instantaneous SNR levels and their corresponding configurations are stored in the lookup table. Then, during actual communication, both the transmitter and receiver quantize the observed SNR value and directly retrieve the appropriate configurations from the lookup table. In Fig. \ref{fig: LookUpTable}, we present the PSNR performances of the JCAMP and JCAP methods using the lookup table across a range of SNR values. The simulation is conducted using the CIFAR-100 dataset and a uniform quantizer that clips the instantaneous SNR values to the range $[0.77, 11.01]$ and quantizes them using 8 bits. This SNR range is determined based on the case where all sub-vectors are assigned to the $V$-th codebook and the first codebook, respectively.
Fig. \ref{fig: LookUpTable} shows that the PSNR of the lookup table-based proposed methods become nearly identical to the unquantized case.

Fig. \ref{fig: codebook_usage} shows the mean of the assigned codebook index for each sub-vector across different SNRs using the JCAMP method over a Rayleigh fading channel with the CIFAR-100 dataset. 
Fig. \ref{fig: codebook_usage} shows that, as the SNR increases, the mean of the assigned codebook index for each sub-vector gradually decreases. This trend occurs because lower-indexed codebooks correspond to smaller $\lambda^{(v)}$ and $\mu_{\rm min}^{(v)}$ values in the proposed multi-codebook training strategy; therefore, these codebooks are associated with lower bit-flip probabilities and offer better task performance. These results confirm that the proposed JCAMP method effectively assigns appropriate codebooks to sub-vectors based on the channel conditions.  

\section{Conclusion}\label{Sec:Conclusion}
In this paper, we proposed ESC-MVQ, a novel digital semantic communication framework that provides a scalable and adaptive solution for diverse communication environments. We developed an end-to-end training strategy that jointly trains multiple VQ codebooks alongside a single semantic encoder-decoder pair. By modeling the transmission and reception of VQ outputs using parallel BSCs with trainable bit-flip probabilities, our strategy enables the trained VQ codebooks to be independent of specific communication scenarios. To maximize inference performance, we devised an optimal communication strategy that jointly optimizes codebook assignment, modulation, and power allocation. Simulation results demonstrated that ESC-MVQ significantly outperforms existing methods, achieving superior PSNR while using only a single encoder-decoder pair, unlike other approaches that require multiple pairs.


\end{document}